\documentclass[aps,prd,twocolumn,superscriptaddress,floatfix,preprintnumbers]{revtex4} 
\usepackage{graphicx}% 
\usepackage{dcolumn}   
\usepackage{bm}        
\usepackage{amsmath}
\usepackage{amssymb}   
\usepackage{color}
\usepackage{ifluatex}
\usepackage[utf8]{\ifluatex lua\fi inputenc}
\usepackage[T1]{fontenc}
\usepackage{epstopdf}
\usepackage{slashed}
\usepackage{lineno}
\usepackage{caption}
\usepackage{hyperref}

\usepackage{tikz}
\usetikzlibrary{positioning,arrows}
\usetikzlibrary{decorations.pathmorphing}
\usetikzlibrary{decorations.markings}
\usetikzlibrary{shapes.geometric}
\usepackage{endnotes}
\tikzset{
    vector/.style={decorate, decoration={snake}, draw},
    provector/.style={decorate, decoration={snake,amplitude=2.5pt}, draw},
    antivector/.style={decorate, decoration={snake,amplitude=-2.5pt}, draw},
    fermion/.style={draw=black,
      postaction={decorate},decoration={markings,mark=at position .55
        with {\arrow[draw=black]{>}}}}, 
    fermionbar/.style={draw=black, postaction={decorate},
                       decoration={markings,mark=at position .55 with {\arrow[draw=black]{<}}}},
    fermionnoarrow/.style={draw=black},
    gluon/.style={decorate, draw=black,decoration={coil,amplitude=2pt, segment length=2.5pt}},
    photon/.style={decorate, draw=black,decoration={snake,amplitude=2pt, segment length=5pt}},
    higgs/.style={dashed,draw=black}, 
    scalar/.style={dashed,draw=black,
      postaction={decorate},decoration={markings,mark=at position .55
        with {\arrow[draw=black]{>}}}}, 
    scalarbar/.style={dashed,draw=black,
      postaction={decorate},decoration={markings,mark=at position .55
        with {\arrow[draw=black]{<}}}}, 
    scalarnoarrow/.style={dashed,draw=black},
    electron/.style={draw=black,
      postaction={decorate},decoration={markings,mark=at position .55
        with {\arrow[draw=black]{>}}}}, 
    bigvector/.style={decorate, decoration={snake,amplitude=4pt}, draw},
} 

\begin{document}
%% definition here
\def\D{{\cal D}}
\def\g{\overline{\cal G}}
\def\gm{\gamma}
\def\M{{\cal M}}
\def\unM{\hat{\cal M}}
\def\unas{ \left( \frac{\hat{a}_s}{\mu_0^{\epsilon}} S_{\epsilon} \right) }
\def\asc{\left( \frac{\hat{\alpha}_s}{4 \pi} \right)}
\def\asr{\left( \frac{\alpha_s}{4 \pi} \right)}
\def\rnM{{\cal M}}
\def\rnas{ \left( a_s  \right) }
\def\b0{\beta_0}
\def\cD{{\cal D}}
\def\cC{{\cal C}}
\def\ct{{\red []}}

\def\G{{\Gamma}}
\def\Gp0{{\Gamma^{'}_0}}
\def\Gp1{{\Gamma^{'}_1}}
\def\Gp2{{\Gamma^{'}_2}}
\def\p{\slashed{p}}
\def\XD{{\cal D}}

\allowdisplaybreaks[4]
% \unitlength1cm

\title{NNLO QCD$\oplus$QED corrections to Higgs production in bottom quark annihilation}
\author{Ajjath A H\footnote{E-mail: {\tt ajjathah@imsc.res.in}}}
\affiliation{The Institute of Mathematical Sciences, HBNI, Taramani, Chennai 600113, India}
\author{Pulak Banerjee\footnote{E-mail: {\tt pulak.banerjee@psi.ch}}} 
\affiliation{Paul Scherrer Institut, CH-5232 Villigen PSI, Switzerland}
\author{Amlan Chakraborty\footnote{E-mail: {\tt amlanchak@imsc.res.in}}}
\affiliation{The Institute of Mathematical Sciences, HBNI, Taramani, Chennai 600113, India}
\author{Prasanna K. Dhani\footnote{E-mail: {\tt prasannakumar.dhani@fi.infn.it}}}
\affiliation{INFN, Sezione di Firenze, I-50019 Sesto Fiorentino, Florence, Italy}
\author{Pooja Mukherjee\footnote{E-mail: {\tt poojamukherjee@imsc.res.in}}}
\affiliation{The Institute of Mathematical Sciences, HBNI, Taramani, Chennai 600113, India}
\author{Narayan Rana\footnote{E-mail: {\tt narayan.rana@mi.infn.it}}}
\affiliation{INFN, Sezione di Milano, Via Celoria 16, I-20133 Milan, Italy}
\author{V. Ravindran\footnote{E-mail: {\tt ravindra@imsc.res.in}}}
\affiliation{The Institute of Mathematical Sciences, HBNI, Taramani, Chennai 600113, India}
% 
% \date{\today}

\preprint{\quad IMSc/2019/05/05, PSI-PR-19-12, TIF-UNIMI-2019-7}

%%%%%%%%%%%%%%%%%%%%%%%%%%%%%%%%%%%%%%%%%%%%%%%%%%%%%%%%%%%%%%
\begin{abstract}
We present next-to-next-to leading order (NNLO) quantum electrodynamics (QED)
corrections to the production of the Higgs boson in bottom quark annihilation 
at the Large Hadron Collider (LHC) in the five flavor scheme.  
We have systematically included the NNLO corrections resulting
from the interference of quantum chromodynamics (QCD) and QED interactions.
We have investigated the infrared (IR) structure of the bottom quark form factor up to 
two loop level in QED and in QCD$\times$QED using K+G equation.   We find that
the IR poles in the form factor 
are controlled by the universal cusp, collinear and soft anomalous dimensions.      
In addition, we derive the QED as well as QCD$\times$QED contributions to soft distribution function
as well as to the ultraviolet renormalization constant of the bottom Yukawa coupling up to
second order in strong coupling and fine structure constant.  Finally, we report our findings on 
the numerical impact of the NNLO results from QED and QCD$\times$QED at the LHC energies
taking into account the dominant NNLO QCD corrections. 
\end{abstract}

\maketitle

\section{Introduction}
\label{sec:intro}

\noindent

The discovery of the Standard Model (SM) Higgs boson by ATLAS \cite{Aad:2012tfa} and CMS 
\cite{Chatrchyan:2012xdj} collaborations at the Large Hadron Collider (LHC) 
has not only put the SM in a strong footing but also opened up a plethora of 
physics programs that can probe physics beyond the SM (BSM).   Since the Higgs boson couples dominantly
to heavy fermions and massive vector bosons,  the corresponding observables are expected to be 
sensitive to new physics.  In order to make definitive claims in the context of BSMs, 
it is henceforth extremely important to understand the Higgs sector of the SM.  
This is possible thanks to dedicated efforts from the LHC collaborations to
measure the properties of the SM Higgs boson to unprecedented accuracy.  Both
ATLAS and CMS collaborations have already measured very precisely the partial width of the Higgs bosons both in the SM 
as well as in several BSM scenarios.  This is the beginning of an era of precision physics at the 
LHC.  These studies will be incomplete without the precise theoretical predictions in the SM as well as BSM.  

At hadron colliders, where underlying scattering events are dominated by strong interaction,
quantum effects are unavoidable.  Quantum Chromodynamics (QCD), the theory of strong interaction, 
plays an important role at the LHC.  
Often, one finds that the leading order predictions from perturbative QCD 
are unreliable due to unphysical scales such as renormalization and factorization scales
and also due to missing higher order radiative corrections.  Radiative corrections from QCD are also large.  
Inclusion of such corrections improves the reliability of the predictions not only by 
making them more precise, but also by reducing the dependency on the unphysical
scales.  

At the LHC, the dominant production channel for the Higgs boson is gluon fusion through
top quark loops \cite{Graudenz:1992pv}.  
Owing to the complexities involved with the two loop massive Feynman integrals,
an effective theory where top quark is integrated out, was proposed to obtain  
the first result at next-to-leading order (NLO) \cite{Dawson:1990zj} for the Higgs boson production.
Later on, in \cite{Djouadi:1991tka,Spira:1995rr}, the NLO corrections, taking into
account the mass of the top quark, were shown to be very close
to the prediction from the effective theory approach \cite{Dawson:1990zj}. 
Thanks to the continued efforts beyond the NLO \cite{Harlander:2002wh, Anastasiou:2002yz, 
Ravindran:2003um,Ravindran:2004mb,Moch:2005ky,Ravindran:2005vv,Ravindran:2006cg}, 
the most precise prediction till date, namely next-to-next-to-next-to
leading order (N$^3$LO) prediction \cite{Anastasiou:2015ema, Mistlberger:2018etf} for the
inclusive production of the Higgs boson in the gluon fusion process, is now available
(See \cite{Ravindran:2006bu,Ahmed:2014uya,Dulat:2018bfe} for rapidity distributions).
In addition, at NNLO accuracy, the tiny effects due to finite top quark mass have already been
computed in \cite{Harlander:2009mq,Pak:2009dg}. Electroweak (EW) corrections 
\cite{Aglietti:2004nj,Actis:2008ug} and mixed QCD-EW corrections  \cite{Anastasiou:2008tj} are 
shown to improve the predictions.

The predictions from the perturbative QCD for the dominant production channel have
reached the level of precision which now requires inclusion of the contributions from the 
sub-dominant channels.  For example, one includes production channels such as
vector boson fusion, associated production with a vector boson, bottom quark annihilation etc.  In addition,
the precise predictions \cite{deFlorian:2016spz} taking into account radiative corrections 
from QCD and EW, are known for many of these processes. 

Among these sub-dominant processes, production of the Higgs boson in bottom quark annihilation
has been a topic of interest both in the SM as well as BSM contexts. 
In the SM, Yukawa couplings of the Higgs boson to the
quarks and leptons are free parameters and precise determination of the couplings is possible at
the LHC.  These couplings are highly sensitive to scales of new physics 
as the mass ($m_h$) of the Higgs boson is close to the EW scale. 
Hence,  both ATLAS \cite{ATLAS-CONF-2019-005} and CMS \cite{Sirunyan:2018koj} collaborations have made
dedicated efforts to measure them precisely.
Among them, bottom Yukawa is one of the most sought one and it is a challenging task for experimentalists. 
Associated production of the Higgs boson with vector bosons 
or with top quarks and its subsequent decay to bottom quarks
have been studied to achieve this. In addition, some interesting proposals can be found in \cite{Englert:2015dlp}.

In the SM, bottom Yukawa coupling is less significant with respect to top Yukawa coupling while in the Minimal Supersymmetric SM (MSSM) \cite{Gunion:1992hs}
the coupling is proportional to $1/\cos\beta$ which can increase the cross section in some
parametric region.  The angle $\beta$ is related to the ratio, denoted
by $\tan\beta$, of the vacuum expectation values of two Higgs doublets.   
The production of Higgs boson(s) in perturbative QCD is studied in 
four flavor and five flavor schemes \cite{Buza:1996wv,Bierenbaum:2009mv,Blumlein:2018jfm}, called 4FS and 5FS, respectively.  
In the former, one assumes that proton sea does not contain bottom quark, 
and they are radiatively generated from gluons in the proton.
These bottom quarks can annihilate to produce the Higgs boson.
Their contributions are enhanced by logarithms of bottom quark mass spoiling the perturbation theory. 
Hence they need to be resummed to obtain reliable predictions.  
In the 5FS, one can avoid these logarithms 
by introducing non-zero bottom quark distributions in the proton.   They are
present due to pair production of bottom quarks from the gluons in the proton sea.  
Since, the leading order contribution in 5FS is two to one, while in 4FS, it is two to three,
computations beyond the leading order are relatively easier in 5FS.
In 4FS, only NLO QCD effects \cite{Dittmaier:2003ej, Dawson:2003kb, Wiesemann:2014ioa}
are known.  On the other hand, in 5FS, NLO \cite{Dicus:1998hs,Balazs:1998sb}, 
NNLO \cite{Harlander:2003ai} and the threshold effects at N$^3$LO
\cite{Ravindran:2006cg,Ahmed:2014cha} (see \cite{Ravindran:2006bu,Ahmed:2014era} for rapidity distributions) are known for sometime.
Also, 5FS cross-section providing the dominant 
cross-section in a  matched prediction \cite{Bonvini:2016fgf,Forte:2015hba} is very well known. 
In \cite{Ebert:2017uel}, resummation of time-like logarithms in SCET framework has been performed.
Recently, for the bottom quark annihilation, complete N$^3$LO corrections \cite{Gehrmann:2014vha, Ahmed:2014pka, Duhr:2019kwi} 
have become available.
In addition, the resummation of threshold contributions \cite{H:2019dcl} 
at N$^3$LO+N$^3$LL accuracy have also been included. 

Unlike the dominant channel, gluon fusion to the Higgs boson, bottom quark annihilation has not received much
attention in the context of EW corrections, presumably because it is already sub-dominant
at the LHC.  In this paper, we make the first attempt to include the QED corrections to the
inclusive production to this channel.  We expect that these corrections could be comparable to
the fixed \cite{Duhr:2019kwi} and resummed \cite{H:2019dcl} results solely 
from third order in perturbative QCD.    

In \cite{deFlorian:2018wcj}, pure QED and mixed QCD$\times$QED contributions have
been obtained for the Drell-Yan (DY) process through Abelianization \cite{deFlorian:2015ujt,deFlorian:2016gvk}
at orders ${\cal O}(\alpha^2)$ and ${\cal O}(\alpha \alpha_s)$, respectively.  
In \cite{deFlorian:2018wcj}, a suitable algorithm is obtained by studying the group 
theory structure of QCD and QED amplitudes that contribute to the partonic 
sub-processes of DY production.  The algorithm contains a set of transformations
on the color factors/Casimirs of SU(N) that transforms QCD results for the partonic sub-processes
to the corresponding QED results.  This way both pure QED as well as QCD$\times$QED contributions
to inclusive production cross section for the Z boson in DY process have been obtained
in \cite{deFlorian:2018wcj} at NNLO level.
Following this approach, we can in principle proceed to obtain
pure QED and mixed QCD$\times$QED contributions to the bottom quark annihilation process 
from the QCD results.  Although the QCD results \cite{Maltoni:2003pn, Harlander:2003ai}
to NNLO are presented for $\text{N}=3$ of SU(N) and hence Abelianization can not be used, 
however, in \cite{Majhi:2010zg}, resonant production of sleptons in a 
R-parity violating supersymmetric model was studied where radiative corrections from 
SU(N) gauge fields with $n_f$ fermions were included to NNLO level.  Since, sleptons couple only to
fermions in this model through Yukawa coupling, these NNLO corrections coincide with the
results of \cite{Maltoni:2003pn, Harlander:2003ai} for $\text{N}=3$.  Hence, we could use the results
given in \cite{Majhi:2010zg} and method of Abelianization to obtain pure QED as well as QCD$\times$ QED
results for bottom quark annihilation to the Higgs boson.  However, in order to
scrutinize the very approach of Abelianization,  we explicitly compute
pure QED and QCD$\times$QED corrections to inclusive production of 
the Higgs boson in bottom quark annihilation up to NNLO level in U(1) and SU(N)$\times$U(1).
In addition, we reproduce the same for the production of Z boson in DY process.
The computation beyond the leading order involves  
evaluation of virtual and real emission processes.  The contributions from them 
are sensitive to ultraviolet (UV) and
infrared (IR) divergences.  We compute them in dimensional regularization, hence divergences appear as
poles in dimensional parameter $\varepsilon = d-4$, $d$ being the space-time dimension.   
The UV divergences are removed in ${\overline{\rm MS}}$  
scheme.  The IR divergences result from soft gluons and massless collinear partons.  The former is called
the soft divergence and later collinear divergence.  While soft divergences cancel between virtual and
real emission processes in the inclusive cross section, the collinear divergences are removed
by mass factorization.  We determine, both UV as well as mass factorization counter terms using 
factorization property of the inclusive cross section and obtain collinear finite 
contributions to the Higgs boson production in bottom quark annihilation and Z boson production in DY.  
We determine IR anomalous dimensions up to two-loop level in both
QED and QCD$\times$QED.  We find that they are process independent.    
Using the universal IR anomalous dimension and following \cite{Ahmed:2015qpa}, 
we compute the renormalization constant
for the Yukawa coupling in QED as well as in QCD$\times$QED from the form factors (FF)
of Higgs bottom anti-bottom operator and vector current of DY process.

The paper is organized as follows. In section \ref{sec:comp}, after discussing 
the theoretical frame work, we briefly describe in sub-section \ref{subsec:method}
how we compute higher order QCD and QED radiative corrections to various partonic and photonic
channels that contribute to the inclusive cross section.  In sub-section \ref{subsec:uv}, we discuss the UV and
IR structure of the form factors and cross sections using K+G equation and obtain the mass factorized
cross sections. In the following sub-section, we discuss about the Abelianization procedure.
The phenomenological impact of our theoretical predictions are presented in section \ref{sec:results}.  
Finally we summarize in section \ref{sec:conclusion}.  
The universal constants that appear in soft distribution
function, FFs of vector current and bottom quarks and the mass factorized partonic and photonic
cross sections are presented in the Appendix \ref{ap:g}, \ref{ap:ff} and \ref{ap:cs}, respectively.

\section{Theoretical framework}
\label{sec:comp}
\noindent
The Lagrangian that describes the interaction of the Higgs boson with the bottom quarks 
is described by the Yukawa interaction and is given by
\begin{equation}
\mathcal{L}_b= -\lambda_{b} \phi(x)\bar{\psi}_{b}(x)\psi_{b}(x)\,,
\end{equation}
where $\lambda_{b}$ is the Yukawa coupling which, after the EW symmetry breaking, 
is found to be ${m_{b}}/{v}$.
$\psi_{b}(x)$ and $m_{b}$ denote the bottom quark field and mass, respectively.
$v$ is the vacuum expectation value ({\it vev}) of the Higgs field $\phi(x)$. 
In the SM, the Higgs boson production through bottom quark annihilation is
sub-dominant compared to gluon fusion through top quark loop.  
One finds that the bottom Yukawa coupling is
35 times smaller than top quark Yukawa coupling and in addition, 
the bottom quark flux in the proton-proton
collision is much smaller than the gluon flux.  
However, in the MSSM \cite{Nilles:1983ge},  $\tan \beta$, the ratio of 
the {\it vev}\,s of Higgs doublets can increase the contributions resulting from the 
bottom quark annihilation channel.  At LO, 
\begin{eqnarray}
\frac{\lambda_{t}^{\rm MSSM}}{\lambda_{b}^{\rm MSSM}}
 = f_{\phi}(\chi) \frac{m_t}{m_b} \frac{1}{\tan \beta} \,,
\end{eqnarray}   
with
\begin{equation}
f_{\phi}(\chi) = 
         \begin{cases} -\cot \chi\,\,\, \text{for}\,\,\, \phi = h,\\ 
                      ~~\tan \chi\,\,\,\text{for}\,\,\, \phi = H, \\ 
                      ~~\cot \beta \,\,\,\,\text{for}\,\,\, \phi = A, 
         \end{cases} 
\end{equation}
where $h$ is the SM like light Higgs boson, $H$ and $A$ are the heavy and the pseudo-scalar Higgs bosons, 
respectively.  The parameter $\chi$ is the mixing angle between weak and mass 
eigenstates of the neutral Higgs bosons $h$ and $H$.  
We set $m_b=0$ except in the Yukawa coupling ~\cite{Aivazis:1993pi, Collins:1998rz, Kramer:2000hn}  
as it is much smaller than the other energy scales in the process. 
The number of active flavors is taken to be $n_f =5$ and we work in the Feynman gauge. 

The inclusive production of a colorless state in hadronic collisions is given by 
\begin{align}
 \sigma (S,q^2) = \sigma_0 (\mu_R^2) &\sum_{cd} \int dx_1 dx_2 
 f_c (x_1, \mu_F^2) f_d (x_2, \mu_F^2) 
 \nonumber\\&
\times \Delta_{cd} (s,q^2, \mu_F^2, \mu_R^2) \,,
\end{align}
where $\sigma_0$ is the Born cross section 
and $f_a (x_i, \mu_F^2)$ are parton distribution functions (PDFs) for $a = q, \overline q ,g$ and 
photon distribution function (PHDF) if $a=\gamma$.  The scaling variables $x_i$ is their momentum fractions.
$\Delta_{cd}$ are the partonic sub-process contributions normalized by the Born cross section.
The scales $\mu_R$ and $\mu_F$ are renormalization and factorization scales. 
$S$ and $s=x_1 x_2 S$ are hadronic and partonic center of mass energy, respectively.
$q^2$ is the invariant mass of the final colorless state.
$\Delta_{cd}$ can be expanded in powers of 
the QCD coupling constant $a_s=g_s^2(\mu_R^2)/16 \pi^2$ and QED coupling constant
$a_e=e^2(\mu_R^2)/16 \pi^2$, $g_s$ and $e$ being the strong and electromagnetic coupling constants, respectively.  That is, 
after suppressing $\mu_R$ and $\mu_F$ dependence,
\begin{equation} \label{eq:delexp}
 \Delta_{cd}(z,q^2) = \sum_{i,j=0}^{\infty} a_s^i a_e^j \Delta_{cd}^{(i,j)}(z,q^2) \,,
\end{equation}
with $\Delta_{cd}^{(0,0)}=\delta(1-z)$ and $z=q^2/s$.
In the following, we describe the methodology to compute $\Delta_{cd}^{(i,j)}$ up to second order
in the couplings.
\subsection{Methodology} \label{subsec:method}
In this section, we briefly describe how higher order perturbative corrections $\Delta_{cd}^{(i,j)}$ 
(Eq. \ref{eq:delexp}) are computed. 
The details of computational procedure can be found in \cite{Ahmed:2016qhu}.
Beyond the leading order (LO), the partonic channels consists of one and two loop virtual sub processes, 
real-virtual and single and double-real emissions, some of which are presented   
in Fig.~\ref{fig:dv}, Fig.~\ref{fig:rv} and Fig.~\ref{fig:dr}.
The black line with an arrow indicates the bottom quark, the wavy line the photon, the curly 
line the gluon and the Higgs boson is indicated by the dashed line.

Sub-processes involving virtual diagrams are sensitive to UV singularities.
Due to the presence of massless gluons and photons, we encounter soft singularities in both virtual and
real emission sub-processes.  In addition, we encounter collinear singularities,
as we treat all the quarks including the bottom quark massless. 
We use dimensional regularization to regulate all these singularities.
\begin{widetext}
\begin{minipage}[b]{0.27\textwidth}
 \centering
\begin{tikzpicture}[scale=0.5]
 \draw[fermion] (-3.0,1.5) -- (0,0);
 \draw[fermion] (0,0) -- (-3.0,-1.5);
 \draw[photon] (-2.0,1.0) -- (-2.0,-1.0);
 \draw[gluon] (-1.0,0.5) -- (-1.0,-0.5);
 \draw[higgs] (0,0) -- (0.7,0);
 \draw[higgs] (1.8,0) -- (2.5,0);
 \draw[fermion] (3.5,-1.3) -- (2.5,0);
 \draw[fermion] (2.5,0) -- (3.5,1.3); 
 \node at (1.25,0) {$\times$};
\end{tikzpicture}
\captionsetup{font=footnotesize}

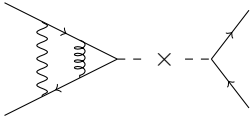
\captionof{figure}{Double virtual contribution}
\label{fig:dv}
\end{minipage}
\begin{minipage}[b]{0.40\textwidth}
 \centering
\begin{tikzpicture}[scale=0.5]
 \draw[fermion] (-2.0,1.3) -- (0,1.3);
 \draw[fermion] (0,1.3) -- (0,-1.3);
 \draw[fermion] (0,-1.3) -- (-2.0,-1.3);
 \draw[gluon] (-1.0,1.3) -- (-1.0,-1.3);
 \draw[photon] (0.0,1.3) -- (0.7,1.3);
 \draw[higgs] (0,-1.3) -- (0.7,-1.3);
 \node at (1.25,1.3) {$\times$};
 \node at (1.25,-1.3) {$\times$}; 
 \draw[higgs] (1.7,-1.3) -- (2.5,-1.3);
 \draw[photon] (1.7,1.3) -- (2.5,1.3); 
 \draw[fermion] (2.5,1.3) -- (3.5,1.3);
 \draw[fermion] (2.5,-1.3) -- (2.5,1.3);
 \draw[fermion] (3.5,-1.3) -- (2.5,-1.3);
\end{tikzpicture}
\quad 
\begin{tikzpicture}[scale=0.5]
 \draw[fermion] (-2.0,1.3) -- (0,1.3);
 \draw[fermion] (0,1.3) -- (0,-1.3);
 \draw[fermion] (0,-1.3) -- (-2.0,-1.3);
 \draw[photon] (-1.0,1.3) -- (-1.0,-1.3);
 \draw[gluon] (0.0,1.3) -- (0.7,1.3);
 \draw[higgs] (0,-1.3) -- (0.7,-1.3);
 \node at (1.25,1.3) {$\times$};
 \node at (1.25,-1.3) {$\times$}; 
 \draw[higgs] (1.7,-1.3) -- (2.5,-1.3);
 \draw[gluon] (1.7,1.3) -- (2.5,1.3); 
 \draw[fermion] (2.5,1.3) -- (3.5,1.3);
 \draw[fermion] (2.5,-1.3) -- (2.5,1.3);
 \draw[fermion] (3.5,-1.3) -- (2.5,-1.3);
\end{tikzpicture}
\captionsetup{font=footnotesize}

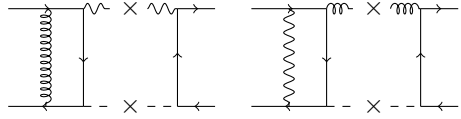
\captionof{figure}{Real virtual contribution}
\label{fig:rv}
\end{minipage}
\begin{minipage}[b]{0.27\textwidth}
 \centering
\begin{tikzpicture}[scale=0.5]
 \draw[fermion] (-1.0,1.3) -- (0,1.3);
 \draw[fermion] (0,1.3) -- (0,-1.3);
 \draw[fermion] (0,-1.3) -- (-1.0,-1.3);
 \draw[photon] (0.0,1.3) -- (1.0,1.3);
 \draw[gluon] (0.0,0) -- (1.0,0);
 \draw[higgs] (0,-1.3) -- (1.0,-1.3);
 \node at (1.45,1.3) {$\times$};
 \node at (1.45,0) {$\times$};
 \node at (1.45,-1.3) {$\times$}; 
 \draw[higgs] (1.9,-1.3) -- (2.9,-1.3);
 \draw[photon] (1.9,1.3) -- (2.9,1.3); 
 \draw[gluon] (1.9,0) -- (2.9,0); 
 \draw[fermion] (2.9,1.3) -- (3.9,1.3);
 \draw[fermion] (2.9,-1.3) -- (2.9,1.3);
 \draw[fermion] (3.9,-1.3) -- (2.9,-1.3);
\end{tikzpicture}
\captionsetup{font=footnotesize}

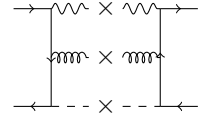
\captionof{figure}{Double real contribution}
\label{fig:dr}
\end{minipage}
\vspace{0.2cm}
\end{widetext}
We have used the program QGRAF \cite{Nogueira:1991ex} to generate virtual as well as real emission
Feynman diagrams that contribute to the relevant sub-processes.
An in-house FORM \cite{Vermaseren:2000nd} code is used 
to perform all the  symbolic manipulations, e.g. performing Dirac, SU(N) color and Lorentz algebra.
Large number of loop integrals show up in the virtual diagrams. 
The integration-by-parts identities are used through a Mathematica based package, LiteRed \cite{Lee:2013mka} to reduce 
them to a minimum set of master integrals.  
For those processes that involve pure real emissions with or without virtual diagrams, we use the
method of reverse unitarity \cite{Anastasiou:2002yz} that allows one
to use IBP identities to reduce the resulting phase-space integrals to
a set of few master integrals, the later can be found in \cite{Anastasiou:2012kq}.
Finally we obtain contributions to each sub-process, containing UV and IR singularities as poles in 
$\varepsilon=d-4$.  

In the next section, we study the UV and IR structure of FF 
and the soft distribution function that contribute to NNLO level in QCD, QED and QCD$\times$QED.  
In order to explore the IR structure, we study the production of a $Z$ boson
in hadron colliders, namely DY process to the
same accuracy in QCD, QED and QCD$\times$QED.  In particular, we focus our
attention to the FF and the soft distribution function that contribute
to the inclusive DY production cross section. 
Following \cite{Ravindran:2006cg, Ahmed:2015qpa, Ahmed:2014cla}, we demonstrate
the factorization of IR singularities in both the FFs and show how to extract the process
independent cusp ($A$), collinear ($B$) and soft ($f$) anomalous dimensions from them.   
Using the FF of the bottom quark and  
the process independent soft distribution function we can extract 
the UV anomalous dimension of the Yukawa coupling
$\lambda_b$ up to two loop level in QCD, QED and QCD$\times$QED.   
Finally, we demonstrate the factorization of collinear singularities and the mass factorization 
leading to IR finite partonic contributions to inclusive hadronic cross sections 
for both the Higgs boson and DY productions up to NNLO level in QCD, QED and QCD$\times$QED.

\subsection{UV and IR structures in QED and \texorpdfstring{QCD$\times$QED}{qcdqed}} \label{subsec:uv}
Having computed all the partonic channels that contribute to the hadronic cross sections in QED and QCD$\times$QED, we use them to study the
UV and IR structure of the FFs and soft gluon/photon emissions in the Higgs boson and DY 
productions.  For the former, we have
used Sudakov K+G equation and for the later, following \cite{Ravindran:2005vv,Ravindran:2006cg} we exploited 
the universal structure of soft distribution function resulting from the soft gluon/photon emissions. 

In order to remove the UV divergences that result from virtual sub-processes, we use  
the renormalization constants
$Z_{a_c}, c=s, e$ for the QCD and QED coupling constants, respectively
and $Z_{\lambda_b}$ for the Yukawa coupling.  
The Yukawa coupling $Z_{\lambda_b}$ receives
contributions from both QCD and QED.
$Z_{a_s}$ and $Z_{a_e}$ relate the bare couplings $\hat{a}_s=\hat g_s^2/16 \pi^2$ of QCD and $\hat{a}_e=\hat e^2/16 \pi^2$ of QED 
to the renormalized ones $a_s(\mu_R^2)$ and $a_e(\mu_R^2)$, respectively, at the renormalization scale $\mu_R$ 
in the following way,
\begin{align} \label{UV1}
\frac{\hat{a}_c}{\big(\mu^2 \big)^{\frac{\varepsilon}{2}}} S_{\varepsilon} = 
\frac{a_c(\mu_R^2)}{\big( \mu_{R}^2 \big)^{\frac{\varepsilon}{2}}} ~ Z_{a_c} \left(a_s(\mu_{R}^2),a_e(\mu_R^2),
\varepsilon\right)\,, 
\end{align}
where $ a_c=\{a_s,a_e\}$.
Here, $S_{\varepsilon} \equiv \exp[\left(\gamma_E-\ln4\pi\right)\frac{\varepsilon}{2}]$
is the phase-space factor in $d$-dimensions, $\gamma_{E} = 0.5772...$ is the Euler-Mascheroni constant and $\mu$ is an arbitrary
mass scale introduced to make $\hat a_s$ and $\hat a_e$ dimensionless in  $d$-dimensions. The renormalization constant $Z_{a_c}$ up to two-loops are given by
\begin{align}
 Z_{a_s} &= 1 + a_s \Big( \frac{2\beta_{00}}{\varepsilon} \Big) 
 + a_s a_e \Big( \frac{\beta_{01}}{\varepsilon} \Big)
 + a_s^2 
\left({4 \beta_{00}^{2} \over \varepsilon^2}+{\beta_{10}\over \varepsilon}\right)
%\nonumber\\&
%+ {\cal O}(a^2_s a_{e}) + {\cal O}(a_s a_{e}^2)
\nonumber\\
 Z_{a_e} &= 1 + a_e \Big( \frac{2\beta'_{00}}{\varepsilon} \Big) 
 + a_e a_s \Big( \frac{\beta'_{10}}{\varepsilon} \Big)
 + a_e^2 
\left({4 \beta_{00}^{'2} \over \varepsilon^2}+{\beta'_{01}\over \varepsilon}\right)
\end{align}
where $\beta_{ij}$ and $\beta'_{ij}$ are QCD and QED beta functions, respectively.
In the present case, only one loop $\beta$ i.e. $\beta_{00}$ and $\beta_{00}'$\cite{Cieri:2018sfk} appear.
They are given by 
\begin{equation}
\beta_{00} = \frac{11}{3} C_A - \frac{4}{3} n_f T_F\,, \quad
\beta_{00}' = -\frac{4}{3} \Big( N \sum_q e_q^2 \Big) \,.
\end{equation}
Here, $C_A=N$ is the adjoint Casimir of $SU(N)$. We also denote the fundamental Casimir $C_F=(N^2-1)/2 N$ for later use. 
$n_f$ is the number of active quark flavors and $e_q$ refers to electric charge for quark $q$.
The renormalization constant $Z_{\lambda}^{b}(a_s, a_e) $ satisfies the renormalization group equation:
\begin{eqnarray}
\mu_R^2 {d \over d\mu_R^2} \ln Z^{b}_\lambda = \frac{\varepsilon}{4} + \gamma^{(i,j)}_{b} (a_s(\mu_R^2),a_e(\mu_R^2))
\end{eqnarray}
whose solution in terms of the anomalous dimensions $\gamma^{(i,j)}_b$ up to two loops is found to be 
\begin{align} \label{Zlam}
Z_{\lambda_b} (& a_s, a_e,\varepsilon ) = 
1  + a_s \Big\{ \frac{1}{\varepsilon}\big(2 \gamma^{(1,0)}_b \big) \Big\} 
   + a_e \Big\{ \frac{1}{\varepsilon}\big(2 \gamma^{(0,1)}_b \big) \Big\}    
\nonumber\\&
   + a_s^2 \Big\{ \frac{1}{\varepsilon^2} \Big( 2 \big(\gamma^{(1,0)}_b \big)^2 + 2 \beta_{00} \gamma^{(1,0)}_b \Big)
                + \frac{1}{\varepsilon} \gamma^{(2,0)}_b \Big\} 
\nonumber \\
  &+ a_e^2 \Big\{ \frac{1}{\varepsilon^2} \Big( 2 \big(\gamma^{(0,1)}_b \big)^2 + 2 \beta_{00}' \gamma^{(0,1)}_b \Big)
                + \frac{1}{\varepsilon} \gamma^{(0,2)}_b \Big\}
\nonumber\\&
   + a_s  a_e \Big\{ \frac{1}{\varepsilon^2}\big(4 \gamma^{(1,0)}_b \gamma^{(0,1)}_b \big) + \frac{1}{\varepsilon}\big(\gamma^{(1,1)}_b \big) \Big\}  \,.
\end{align}
Note that while the UV singularities factorize through $Z_{\lambda_b}$, singularities from QCD and QED
mix from two loop onward.
For QCD, $\gamma^{(i,0)}_b$ is known to four loops \cite{Vermaseren:1997fq}.  In this paper, 
using the universal IR structure of the amplitudes and cross sections in QED, 
we determine $\gamma^{(i,j)}_b$ up to two loops in QED i.e. for $(i,j) = (0,1),(0,2)$ and in QCD$\times$QED i.e.
for $(i,j) = (1,1)$.  

We begin with the bare form factors $\hat{F}_I(\hat{a}_s,\hat{a}_e,Q^2,\mu^2)$, $I=q,b$ 
where $q$ denotes the DY process and $b$ denotes the Higgs boson production in bottom quark annihilation.  
Note that, these FFs are computed in the perturbative framework where both QCD as well as QED interactions
are taken into account simultaneously and hence they depend on both QCD and QED coupling constants.
In addition, we find that the UV renormalized FFs demonstrate the factorization of IR singularities.  
Using gauge and renormalization group invariance, we propose Sudakov integro-differential equation for these FFs, analogous
to the QCD one.  In dimensional regularization, they take the following form 
\begin{align}\label{eq:KG}
Q^2\frac{d}{dQ^2}\ln \hat{F}_I
= \frac{1}{2}\Big[&  K_I \Big( \{\hat{a}_c\}, \frac{\mu_R^2}{\mu^2}, \varepsilon \Big) 
\nonumber \\&
+ G_I \Big( \{\hat{a}_c\}, \frac{Q^2}{\mu_R^2},\frac{\mu_R^2}{\mu^2},\varepsilon \Big) \Big]
\,,
\end{align}
where $\{a_c\} = \{a_s, a_e \}$ and $Q^2=-q^2$ is the invariant mass of the final state particle
(di-lepton pair in the case of DY and single Higgs boson for the case of Higgs production).
Explicit computation of the form factors shows that IR singularities, resulting from QCD and QED interactions 
not only factorize but also mix beyond one loop level.  In other words, if we factorize IR singularities from the
FFs, the resulting IR singular function can not be written as a product of pure QCD and pure 
QED functions. 
More specifically, there will be terms proportional to $a_s^i a_e^j$, where $i,j > 0$, 
which will not allow factorization of 
QCD and QED ones.  Hence, $K_I$ will have IR poles in $\varepsilon$ from pure QED and pure QCD in every order in
perturbation theory and in addition, from 
QCD$\times$QED starting from ${\cal O} (a_s a_e)$.
On the other hand, overall factorization of IR singularities implies that 
the constants $K_I$ contain the IR singularities from QCD, QED and QCD$\times$QED, 
while the $G_I$s will have IR finite contributions.
Since, the IR singularities of FFs have dipole structure, $K_I$ will be independent of $q^2$ while 
$G_I$s will be finite in $\varepsilon \rightarrow 0$ and the later contain only logarithms in $q^2$.  
Note that $\hat{F}_I$ are renormalization group (RG) invariant so does the sum $K_I+G_I$.
Thus, the RG invariance of  $\hat{F}_I$ implies 
\begin{eqnarray}
\mu_R^2\frac{d}{d\mu_R^2}K_I \Big( \{\hat{a}_c\},
\frac{\mu_R^2}{\mu^2},\varepsilon \Big) = -A_I(\{a_c(\mu_R^2)\} 
) \,,
\label{eq:KSoln} \\
\mu_R^2\frac{d}{d\mu_R^2}G_I \Big( \{\hat{a}_c\},
\frac{Q^2}{\mu_R^2},\frac{\mu_R^2}{\mu^2},\varepsilon \Big) = A_I(\{a_c(\mu_R^2)\} 
%,a_e(\mu_R^2)
)\,,
\label{eq:GSoln} 
\end{eqnarray}   
where $A_I$ are the cusp anomalous dimensions.
The solutions to the above RG equations for $K_I$ can be obtained by expanding the 
cusp anomalous dimensions ($A_I$) in powers of renormalized coupling constants $a_s(\mu_R^2)$ and  
$a_e(\mu_R^2)$ as 
\begin{eqnarray}
A_I (\{a_c(\mu_R^2)\}) =  \sum_{i,j} a_s^i(\mu_R^2) a_e^j(\mu_R^2) A_I^{(i,j)}
\end{eqnarray}
and $K_I$ as
\begin{eqnarray}
\label{Kexp}
K_I(\mu_R^2,\varepsilon)  =  \sum_{i,j}\hat{a}_s^i \hat{a}_e^j \Big( 
\frac{\mu_R^2}{\mu^2}\Big)^{(i+j) \frac{\varepsilon}{2}}S_{\varepsilon}^{(i+j)} K_I^{(i,j)}(\varepsilon) \,,
\end{eqnarray}
where $A^{(i,0)}$ and $A^{(0,i)}$ result from pure QCD and pure QED interactions and $A^{(i,j)}, i,j>0$ from 
QCD$\times$QED.   Using RG equations for the couplings  $a_s$ and $a_e$,
the perturbative solutions to Eq.(\ref{eq:KSoln}) are found to be,
\begin{align}\label{eq:KFF}
K_I^{(1, 0)} &= \frac{1}{\varepsilon}\Big(-2A_I^{(1, 0)}\Big) \,.
\nonumber\\ 
K_I^{(0, 1)} &= \frac{1}{\varepsilon}\Big(-2A_I^{(0, 1)}\Big) \,.
\nonumber\\ 
K_I^{(2, 0)} &= \frac{1}{\varepsilon^2}\Big(2\beta_{00} A_I^{(1, 0)}\Big) + \frac{1}{\varepsilon}\Big( -A_I^{(2,0)}\Big) \,.
\nonumber\\
K_I^{(0, 2)} &= \frac{1}{\varepsilon^2}\Big(2\beta_{00}' A_I^{(0, 1)} \Big) + \frac{1}{\varepsilon}\Big( -A_I^{(0, 2)}\Big) \,.
\nonumber\\
K_I^{(1, 1)} &= \frac{1}{\varepsilon}\Big(-A_I^{(1, 1)}\Big) \,.
\end{align}
Unlike $K_I$, $G_I$ do not contain any IR singularities but depend only on $Q^2$ and hence we expand them as  
\begin{align}
G_I \Big( \{\hat{a_c}\}, \frac{Q^2}{\mu_R^2},\frac{\mu_R^2}{\mu^2},\varepsilon \Big) 
&= G_I(\{a_c(Q^2)\},1,\varepsilon) 
\nonumber \\
&+ \int_{\frac{Q^2}{\mu_R^2}}^{1} \frac{d\lambda^2}{\lambda^2} A_I(\{a_c(\lambda^2\mu_R^2)\}
) 
\end{align}
where the first term is the boundary condition on each $G_I$ at $\mu_R^2=Q^2$.
Expanding $A_I$ in powers of $a_s$ and $a_e$ 
and using RG equations for QCD and QED couplings, we obtain 
\begin{align}
 \int_{\frac{Q^2}{\mu_R^2}}^{1} \frac{d\lambda^2}{\lambda^2} & A_I(\{a_c(\lambda^2\mu_R^2)\}) 
= \sum_{i,j} \hat{a}_s^i \hat{a}_e^j \Big( \frac{\mu_R^2}{\mu^2} \Big)^{(i+j) \frac{\varepsilon}{2}} 
\nonumber\\&
\times S_{\varepsilon}^{(i+j)} \Big[ \Big( \frac{Q^2}{\mu_R^2} \Big)^{(i+j) \frac{\varepsilon}{2}} - 1 \Big] K^{(i,j)}(\varepsilon) \,.
\end{align}
Expanding the finite function $G_I(a_s(Q^2),a_e(Q^2),1,\varepsilon)$ as,
\begin{eqnarray}\label{Gexp}
G_I( \{a_c(Q^2)\}, 1, \varepsilon ) = \sum_{i,j}a_s^i(Q^2)a_e^j(Q^2) G_I^{(i,j)}(\varepsilon) \,,
\end{eqnarray}
substituting the solutions of $K_I$ and $G_I$ in Eq.(\ref{eq:KG}) and performing the integration over $Q^2$ we get
\begin{eqnarray}
\ln \hat{F}_I
= \sum_{i,j}\hat{a}_s^i \hat{a}_e^j\Big( \frac{Q^2}{\mu^2} \Big)^{(i+j)\frac{\varepsilon}{2}}S_{\varepsilon}^{(i+j)}\hat {\mathcal{L}}_{F_I}^{(i,j)}(\varepsilon) \,,
\end{eqnarray}
where,
\begin{align} \label{LFo}
\hat {\mathcal{L}}_{F_I}^{(1,0)} &= \frac{1}{\varepsilon^2}\Big( -2A_I^{(1,0)} \Big) + \frac{1}{\varepsilon}\Big( G_I^{(1,0)}(\varepsilon) \Big) \,. 
\nonumber \\
\hat {\mathcal{L}}_{F_I}^{(0,1)} &= \frac{1}{\varepsilon^2}\Big( -2A_I^{(0,1)} \Big) + \frac{1}{\varepsilon}\Big( G_I^{(0,1)}(\varepsilon) \Big) \,.
\nonumber\\
\hat {\mathcal{L}}_{F_I}^{(2,0)} &= \frac{1}{\varepsilon^3}\Big( \beta_{00} A_I^{(1,0)} \Big) 
           + \frac{1}{\varepsilon^2}\Big( -\frac{1}{2}A_I^{(2,0)} 
\nonumber\\&           
           - \beta_{00} G_I^{(1,0)}(\varepsilon) \Big)
           + \frac{1}{2\varepsilon}\Big( G_I^{(2,0)}(\varepsilon) \Big) \,.
\nonumber\\
\hat {\mathcal{L}}_{F_I}^{(0,2)} &= \frac{1}{\varepsilon^3}\Big( \beta_{00}' A_I^{(0,1)} \Big) 
           + \frac{1}{\varepsilon^2}\Big( -\frac{1}{2}A_I^{(0,2)} 
\nonumber\\&           
           - \beta_{00}' G_I^{(0,1)}(\varepsilon) \Big)
           + \frac{1}{2\varepsilon}\Big( G_I^{(0,2)}(\varepsilon) \Big) \,.
\nonumber \\
\hat {\mathcal{L}}_{F_I}^{(1,1)} &= \frac{1}{\varepsilon^2}\Big( -\frac{1}{2}A_I^{(1,1)} \Big) + \frac{1}{2\varepsilon}\Big( G_I^{(1,1)}(\varepsilon) \Big) \,.
\end{align}
Following \cite{Ravindran:2004mb},  we expand $G_I^{(i,j)}(\varepsilon)$ around $\varepsilon=0$ in terms of  
collinear ($B_I^{(i,j)}$), soft ($f_I^{(i,j)}$) and UV ($\gamma_I^{(i,j)}$) anomalous dimensions as
\begin{align} \label{G}
G_I^{(i,j)}(\varepsilon) &= 2(B_I^{(i,j)} - \gamma_I^{(i,j)} ) + f_I^{(i,j)} +  \sum_{k=0}\varepsilon^k g_{I,ij}^{k} 
\end{align}
with
\begin{align}
 g_{I,10}^0 &= 0 \,, \quad  g_{I,01}^0 = 0 \,, \quad g_{I,11}^0 = 0 \,,
\nonumber\\
 g_{I,20}^0 &= - 2 \beta_{00} g_{I,10}^1 \,, \quad  g_{I,02}^0 = - 2 \beta_{00}' g_{I,01}^1 \,.
\end{align}
The form factors $\hat F_I$ that we computed in this paper in  
QCD, QED and QCD$\times$QED up to two loop level can be used to extract
the cusp anomalous dimensions ($A_I^{(i,j)}$) by comparing them against Eq.(\ref{LFo}).  
We find $A_I^{(i,j)}$ up to two loops as
\begin{align}
A_I^{(1,0)} &= 4 C_F \,. \nonumber \\
A_I^{(0,1)} &= 4 e_I^2 \,. \nonumber \\
A_I^{(2,0)} &= 8 C_A C_F\Big(\frac{67}{18} - \zeta_2 \Big) + 8C_Fn_fT_F\Big(-\frac{10}{9}\Big) \,. \nonumber \\
A_I^{(0,2)} &= 8e_I^2 \Big( N\sum_{k =1}^{n_f}e_{k}^2 \Big) \Big(- \frac{10}{9} \Big) \,. \nonumber \\
A_I^{(1,1)} &= 0 \,.
\end{align}
Unlike $A_I^{(i,j)}$, the 
other anomalous dimensions $B_I^{(i,j)}$, $f_I^{(i,j)}$ and $\gamma^{(i,j)}_I$ 
($\gamma^{(i,j)}_q$ is zero) can not be disentangled either from $\hat F_q$ or $\hat F_b$ alone.  
In order to disentangle $B_I^{(i,j)}$ and $f_I^{(i,j)}$,  we study the partonic cross sections
resulting from soft gluon and soft photon emissions as they are only sensitive to $f_I^{(i,j)}$.

%%%%%%%%%%%%%%%%%%%%%%%%%%%%%%%%%%%%%%%%%%%%%%%%%%%%%%%%%%%%%%%%%%%%%%%
To obtain the process independent part of 
soft gluon/photon contributions in the real emission sub-processes, 
we follow the method described in \cite{Ravindran:2005vv,Ravindran:2006cg},  
where the soft distribution function for the inclusive cross section for producing a colorless state
was obtained from the form factors and partonic sub-process cross sections involving real emissions of gluons.
The soft distribution functions denoted by 
$\Phi_J$, are governed by cusp ($A_J$) and soft anomalous dimensions $f_J$, where
$J=q,b,g$.  It is also known that the identity $\Phi_b=\Phi_q=C_F/C_A \Phi_g$ holds up to
three loop level \cite{Ravindran:2005vv,Ravindran:2006cg,Ahmed:2014cla}.
We can use the partonic sub-processes of either DY process or the
Higgs boson production in bottom quark annihilation  
namely $\hat \sigma_{q \overline q}$ or $\hat \sigma_{b \overline b}$ normalized by the square of the 
bare form factor $\hat F_q$ or $\hat F_b$ to obtain $\Phi_I$. In general $\Phi_I$, which is function of the scaling variable $z = q^2/s$, is defined as,
\begin{equation}
\label{phidef}
\mathcal{C} \exp\Big( 2 \Phi_{I}(z)\Big)=
{\hat \sigma_{I\overline I}(z)
\over
Z_I^2 \big| \hat F_I \big|^2 }  \quad\quad I = q,b
\end{equation}
with $Z_q=1$ and $Z_b=Z_{\lambda_b}$ being the overall renormalization constant. 
The symbol $\mathcal{C}$ refers to ``ordered exponential'' which has the following expansion:
\begin{eqnarray}
\mathcal{C}e^{f(z)} = \delta(1-z) + \frac{1}{1!}f(z) + \frac{1}{2!}(f\otimes f)(z) + \cdots
\end{eqnarray}
Here $\otimes$ is the Mellin convolution and $f(z)$ is a distribution of the kind $\delta(1-z)$ and $\mathcal{D}_i$. The plus distribution $\mathcal{D}_i$ is defined as,
\begin{equation}
 \mathcal{D}_i = \Bigg( \frac{\ln^i (1-z)}{(1-z)} \Bigg)_+ \,.
\end{equation}
We can compute the UV finite 
$\hat \sigma_{ I \overline I}$ every order in renormalized perturbation theory.  Since,
we have not determined $Z_{\lambda_{b}}$, we can only compute the unrenormalized partonic cross section 
$\tilde \sigma_{I \overline I} =\hat \sigma_{I \overline I}/Z_I^2$. 
From the explicit results for $\tilde \sigma_{I \overline I}$ and the form factors $\hat F_I$, using Eq.
(\ref{phidef}) we 
obtain $\Phi_I$ up to second order in $a_s$, $a_e$ and $a_s a_e$.   
We find $\Phi_q=\Phi_b$ up to second order in the couplings demonstrating the universality.
In \cite{Ravindran:2005vv,Ravindran:2006cg}, it was shown
that the soft distribution function $\Phi_I$ satisfies Sukakov K+G equation analogous to the form 
factor $\hat F_I$ due to similar IR structures that both of them have, order by order in perturbation
theory.  That is, $\Phi_I$ satisfies   
\begin{align}\label{eq:KGphi}
q^2\frac{d}{dq^2}\Phi_I = \frac{1}{2} \Big[ \overline{K}_I \Big(& \{\hat{a}_c\}, \frac{\mu_R^2}{\mu^2},\varepsilon,z \Big) 
\nonumber\\
&
+ \overline{G}_I \Big( \{\hat{a}_c\},\frac{q^2}{\mu_R^2},\frac{\mu_R^2}{\mu^2},\varepsilon,z \Big) \Big] \,,
\end{align}
where, the IR singularities are contained in $\overline{K}$ and the finite part in $\overline{G}$. 
RG invariance of $\Phi_I$ implies
\begin{align}
\mu_R^2\frac{d}{d\mu_R^2}\overline{K}_I &= A_I(\{a_c(\mu_R^2)\})\delta(1-z) \,,
\nonumber\\
\mu_R^2\frac{d}{d\mu_R^2}\overline{G}_I &= -A_I(\{a_c(\mu_R^2)\})\delta(1-z) \,.
\end{align}
Note that, the same anomalous dimensions govern the evolution of both $\overline K_I$ and $\overline G_I$.
This ensures that the soft distribution function contains right soft singularities to 
cancel those from the form factor leaving bare partonic cross section to contain only initial
state collinear singularities.  The later will be removed by mass factorization by appropriate
Altarelli-Parisi kernels.   Expanding $\overline K_I(\{a_c\})$ and $\overline{G}_I(\{a_c(q^2)\}, 
1,\varepsilon,z)$ in powers of $\{a_c\}$ as has been done for $K_I(\{a_c\})$ and $G_I (\{a_c\})$ (see Eqs.(\ref{Kexp},\ref{Gexp})),
with the replacements of $K_I^{(i,j)}$ by $\overline {K}_I^{(i,j)}$ and  
\begin{eqnarray}
\label{eq:Gbar}
\overline{G}_I(\{a_c(q^2)\}
,1,\varepsilon,z) = \sum_{i,j}a_s^i(q^2)a_e^j(q^2)\overline{G}_I^{(i,j)}(\varepsilon,z) \,,
\end{eqnarray}
the solution to Eq.(\ref{eq:KGphi}) is found to be
\begin{align}\label{Phi}
\Phi_I(\{\hat{a}_c\},
q^2,\mu^2, &\varepsilon,z) = \sum_{i,j}\hat{a}_s^i\hat{a}_e^j\Big(\frac{q^2(1-z)^2}{\mu^2}\Big)^{(i+j)\frac{\varepsilon}{2}}
\nonumber\\
&
\times S_{\varepsilon}^{(i+j)}
\Big(\frac{(i+j)\varepsilon}{1-z}\Big)\hat{\phi}_I^{(i,j)}(\varepsilon)  
\end{align} 
where,
\begin{eqnarray}
\hat{\phi}_I^{(i,j)}(\varepsilon) = \frac{1}{(i+j)\varepsilon}\Big[ \overline{K}_I^{(i,j)}(\varepsilon) + \overline{G}_I^{(i,j)}(\varepsilon) \Big]\,.
\end{eqnarray} 
$\overline{G}_I^{(i,j)}(\varepsilon)$ is related to finite 
function $\overline{G}_I(\{a_c(q^2)\},1,\varepsilon,z)$ 
defined in Eq.(\ref{eq:Gbar}) through the distributions $\delta(1-z)$ and $\mathcal{D}_j$. 
Thus expanding $\overline{G}^{(i,j)}(\varepsilon)$ in terms of the $a_s(q^2(1-z)^2)$ and $a_e(q^2(1-z)^2)$ we write,
\begin{align}
\sum_{i,j}\hat{a}_s^i\hat{a}_e^j \Big(\frac{q_z^2}{\mu^2}\Big)^{(i+j) \frac{\varepsilon}{2}} & S_\varepsilon^{(i+j)} \overline{G}_I^{(i,j)}(\varepsilon) 
\nonumber\\
&= \sum_{i,j}a_s^i\big(q_z^2\big) a_e^j\big(q_z^2 \big) \overline{\mathcal{G}}_I^{(i,j)}(\varepsilon)
\end{align}
where $q_z^2 = q^2(1-z)^2$.
Following, \cite{Ravindran:2005vv,Ravindran:2006cg}, the IR finite 
$\overline{\mathcal{G}}^{(i,j)}_I (\varepsilon)$ can be expanded as
\begin{align}
\mathcal{\overline G}_I^{(i,j)}(\varepsilon) &= -f_I^{(i,j)} + \sum_{k=0}\varepsilon^k \mathcal{\overline G}_{I,ij}^{(k)} 
\end{align}
where, for up to two loops
\begin{align}
&\mathcal{\overline G}_{I,10}^{(0)} = 0\,, 
 \quad
 \mathcal{\overline G}_{I,01}^{(0)} = 0\,,
 \quad
 \mathcal{\overline G}_{I,11}^{(0)} = 0\,,
 \nonumber\\&
 \mathcal{\overline G}_{I,20}^{(0)} = -2 \beta_{00} \mathcal{\overline G}_{I,10}^{(1)}\,,
 \quad
 \mathcal{\overline G}_{I,10}^{(0)} = -2 \beta_{00}' \mathcal{\overline G}_{I,01}^{(1)} \,.
\end{align}
Comparing the soft distribution functions $\Phi_I$, $I=q,b$, obtained from the explicit computation up to 
second order in coupling constants against the formal solution given in Eqs.(\ref{Phi}), we can
obtain $A_I^{(i,j)}$ and $f_I^{(i,j)}$ for $(i,j)=(1,0),(0,1),(1,1),(2,0),(0,2)$.  
Finally, we obtain $ f_I^{(1,0)}=f_I^{(0,1)}=f_I^{(1,1)} = 0$  and
\begin{align}
\label{f}
f_I^{(2,0)} &= C_A C_F\Big(-\frac{22}{3}\zeta_2  -28 \zeta_3 + \frac{808}{27} \Big) 
\nonumber \\
&
+ C_Fn_fT_F\Big(\frac{8}{3}\zeta_2 - \frac{224}{27}\Big) \,, \nonumber \\
f_I^{(0,2)} &=  e_I^2 \Big( N\sum_{q} e_{q}^2 \Big) \Big(\frac{8}{3}\zeta_2 - \frac{224}{27}\Big) \ . 
\end{align}
Now that we have $f_I^{(i,j)}$, it is now straightforward to obtain $B_q^{(i,j)}$ in Eq.(\ref{G}) from 
the explicit results on $G^{(i,j)}_q$ as
$\gamma^{(i,j)}_q = 0$ for DY.  This way we obtain,
\begin{align} \label{B}
B^{(1,0)}_q &= 3C_F \,, \qquad
B^{(0,1)}_q = 3 e_q^2 \,,
\nonumber \\
B_q^{(2,0)} &= \frac{1}{2} \Big \{ C_F^2 \big (3-24\zeta_2+48\zeta_3 \big) + C_AC_F\Big(\frac{17}{3} + \frac{88}{3}\zeta_2 
\nonumber\\&
-24 \zeta_3  \Big) + C_Fn_fT_F\Big(-\frac{4}{3} -\frac{32}{3}\zeta_2 \Big) \Big\} \,, \nonumber \\
B_q^{(0,2)} &= \frac{1}{2} \Big \{ e_q^4 \big (3-24\zeta_2+48\zeta_3 \big) 
\nonumber\\&
+ e_q^2 \Big( N \sum_{q'} e_{q'}^2\Big) \Big(-\frac{4}{3} -\frac{32}{3}\zeta_2 \Big) \Big\} \,,
\nonumber\\
B_q^{(1,1)} & = C_F e_q^2 \Big (3-24\zeta_2+48\zeta_3 \Big) \,.       
\end{align}
Assuming $B_b^{(i,j)} = B_q^{(i,j)}$, we determine the UV anomalous dimension, $\gamma_b^{(i,j)}$ 
from $G^{(i,j)}_b$ (Eq.(\ref{G})) which is known to second order. 
They are found to be  
\begin{align}
 \gamma^{(1,0)}_b &= 3 C_F \,, \nonumber \\
 \gamma^{(0,1)}_b &= 3 e_b^2 \,, \nonumber \\
 \gamma^{(1,1)}_b & = 3 C_F e_b^2 \,,       \nonumber \\ 
 \gamma^{(2,0)}_b &=    \frac{3}{2} C_F^2 + \frac{97}{6} C_AC_F -\frac{10}{3} C_F n_f T_F \,, \nonumber \\
 \gamma^{(0,2)}_b &=    \frac{3}{2} e_b^4 - \frac{10}{3} e_b^2 \Big( N\sum_{k \in Q}e_{k}^2\Big) \,. 
\end{align}
Alternatively, assuming $B_b^{(i,j)} =B_q^{(i,j)}$ and $f_b^{(i,j)}=f_q^{(i,j)}$,
we can determine $\gamma^{(i,j)}_b$ by comparing  the difference $G_b^{(i,j)}-G_q^{(i,j)}$  
obtained using  DY and Higgs boson form factors $\hat F_q$ and $\hat F_b$ at
$\varepsilon=0$ against the formal decomposition of $G_I^{(i,j)}$ given in  Eqs.(\ref{G}).  
Substituting the above UV anomalous dimensions in Eq.(\ref{Zlam}), we obtain 
$Z_{\lambda_b}$ to second order in the couplings.

Using the renormalization constants $Z_{a_s}$, $Z_{a_e}$ and $Z_{\lambda_b}$ 
for the coupling constants $\alpha_s$,
$\alpha_e$ and the Yukawa coupling $\lambda_b$, we obtain UV finite partonic cross sections. 
The soft and collinear singularities arising from gluons/photons/fermions in the virtual sub-processes 
cancel against those from the real sub-processes
when all the degenerate states are summed up, thanks to the KLN theorem \cite{Kinoshita:1962ur, Lee:1964is}.
What remains at the end, is the initial state collinear singularity, which can be removed by 
mass factorization. Collinear factorization allows us to determine the mass factorization
kernels $\Gamma_{qq}$ and $\Gamma_{qg}$ up to
two-loop level for $U(1)$ and $SU(N)$ $\times$ $U(1)$ cases.  
Since $\Gamma_{qq}$ and $\Gamma_{qg}$ 
are governed by the splitting functions $P_{qq}$ and $P_{qg}$, we extract them to second order in couplings.
In \cite{deFlorian:2015ujt}, these splitting functions up to NNLO level, both in QED and QCD$\times$QED,
were obtained using the Abelianization procedure.
The splitting functions that we have obtained by demanding finite-ness of the
mass factorised cross section, agree with those in \cite{deFlorian:2015ujt}. 
The mass factorized partonic cross section for each partonic sub-process
up to NNLO in QED and in QCD$\times$QED are presented in the Appendix along with 
the known NNLO QCD results \cite{Harlander:2003ai}.   In the next section, we use them to
study their numerical impact at the LHC energies.

\subsection{Abelianization procedure}

In \cite{deFlorian:2018wcj}, QCD$\times$QED corrections to the DY process were 
obtained by studying the $SU(N)$ color factors in Feynman diagrams that contribute to QCD corrections.  
This led to an algorithm namely  Abelianization procedure which provides a set of rules
that transform QCD results into pure QED and mixed QCD$\times$QED results.  
Unlike in \cite{deFlorian:2018wcj}, without resorting to Abelianization rules, 
we have performed explicit calculation to obtain the contributions resulting from all the partonic 
and photonic channels taking into
account both UV and mass factorization counter terms. 
Using these results at NNLO in QCD, QCD$\times$QED and in QED, 
we find a set of rules that can relate QCD and QED results.
Note that if there is a gluon in the initial state,
averaging over its color factor gives a factor $\frac{1}{N^2-1}$.  
This is absent for the processes where photon is present instead of gluon in the initial state.   
Also, for pure QCD or QED, the gluons or photons are degenerate and hence one needs to account 
for a factor of 2.  Keeping these in mind, we arrive at a set of relations among QCD and QED results. 
We have listed them in the following tables for various scattering channels.
They are found to be consistent with the procedure used in \cite{deFlorian:2018wcj}.

{\textbf{Rule 1} : {\textit{quark-quark initiated cases}
\begin{center}
\begin{tabular}{ c | c | c }
%  \hline 
 \quad QCD \quad & \quad QCD$\times$QED \quad & \quad QED \quad \\
 \hline 
%  \midrule
 $C_F^2$ & $2 C_F e_b^2$ & $e_b^4$ \\
 \hline 
 $C_F C_A$ & 0 & 0 \\
 \hline
 $C_F n_f T_F$ & 0 & $e_b^2 \left( N \sum_q e_q^2 \right)$ \\
 \hline 
 $C_F T_F$ & 0 & $N e_b^2 e_q^{2*}$ \\
 \hline 
\end{tabular}
\end{center}
$^{*}e_q^2=e_b^2$ when both initial quarks are bottom quarks.

{\textbf{Rule 2} : {\textit{quark-gluon initiated cases}\\
{(After multiplying $2 C_A C_F$ for the initial state gluon)}
\begin{center}
\begin{tabular}{ c | c | c }
%  \hline 
 \quad QCD \quad & \quad QCD$\times$QED \quad & \quad QED \quad \\
 \hline 
%  \midrule
 $C_A C_F^2$ & $C_A C_F e_b^2$ & $C_A e_b^4$ \\
 \hline 
 $C_A^2 C_F$ & 0 & 0 \\
 \hline
\end{tabular}
\end{center}

{\textbf{Rule 3} : {\textit{gluon-gluon initiated cases}\\
{\small (After multiplying $2 C_A C_F$ for each initial state gluon)}
\begin{center}
\begin{tabular}{ c | c | c }
%  \hline 
 \quad QCD \quad & \quad QCD$\times$QED \quad & \quad QED \quad \\
 \hline 
%  \midrule
 $C_A^2 C_F^2$ & $C_A^2 C_F e_b^2$ & $C_A^2 e_b^4$ \\
 \hline 
 $C_A^3 C_F$ & 0 & 0 \\
 \hline
\end{tabular}
\end{center}

\section{Results and phenomenology} 
\label{sec:results}
\noindent
In this section, we study the numerical impact of pure QED and mixed QCD$\times$QED corrections over the
dominant QCD corrections up to NNLO level to the production of the Higgs boson in bottom quark annihilation at the LHC, mainly for the center 
of mass (CM) energy of $\sqrt{S}=13$~TeV.  
Since we include QED effects, we need PHDF inside the proton
in addition to the standard PDFs. 
For this purpose, we use NNPDF 3.1 LUXqed set \cite{Bertone:2017bme}, 
MRST \cite{Martin:2004dh}, CT14 \cite{Schmidt:2015zda} and PDF4LHC17. 
The PDFs, PHDFs and the strong coupling constant $a_s$ can be obtained, 
using the LHAPDF-6 \cite{Buckley:2014ana} interface.  We have used the following 
input parameters for the masses and the couplings:
\begin{center}
\begin{tabular}{c l r l}
%  \hline
 $m_W$ &= 80.4260~GeV & \quad $m_b (m_b)$ &= 4.70~GeV \\ 
 $m_Z$ &= 91.1876~GeV & $\alpha_{\rm s} (m_h)$ &= 0.113 \\
 $m_h$ &= 125.09~GeV & $\alpha_e$ &= 1/128.0 \\
\end{tabular}
\end{center}
Both $a_s (\mu_R)$ and $m_b (\mu_R)$ are evolved using appropriate QCD $\beta$-function coefficients and quark mass anomalous dimensions 
respectively. However, we have considered fixed $\alpha_e = 4 \pi a_e$ throughout the computation.\\
\begin{figure}[h!]
\includegraphics[width=0.45\textwidth]{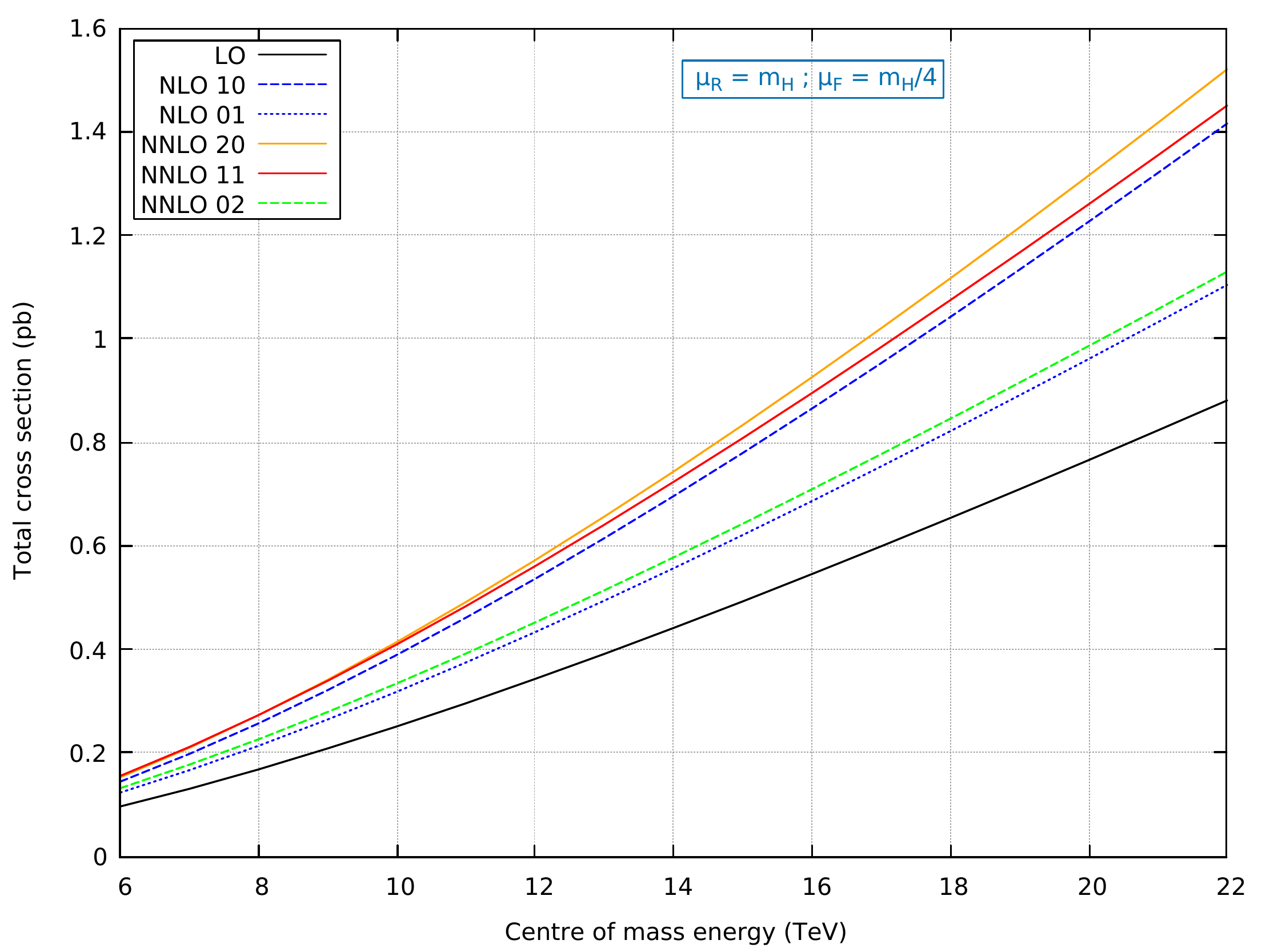}
\caption{\sf The total cross section at various perturbative orders at energy scales varying from 6 to 22 TeV at LHC.}
\label{fig:csrsvar}
\end{figure}
The Higgs boson production cross section from bottom quark annihilation at the present energy of LHC is not substantial.
However, for the high luminosity LHC, measuring them
at higher center of mass energy (CM) would give larger contributions and it will improve the precision.
Hence, we have first studied how the cross section varies with the CM of LHC.   In Fig.~\ref{fig:csrsvar}, 
we plot the inclusive production cross sections 
at various orders in perturbative QCD and QED for the range of CM energies 
between $\sqrt{S}=6$ to $22$~TeV.  In the inset, the index `ij' indicates that QCD at `i'-$th$ order  and QED 
at `j'-$th$ order in perturbative theory are included (\textit{e.g.} `NNLO 11' indicates NNLO mixed QCD$\times$QED).
In Fig.~\ref{fig:csrsvar}, we have used {\small \sf NNPDF31\_lo\_as\_0118, NNPDF31\_nlo\_as\_0118\_luxqed} 
and {\small \sf NNPDF31\_nnlo\_as\_0118\_luxqed} for LO, NLO and NNLO, respectively. 
The renormalization ($\mu_R$) and factorization ($\mu_F$) scales are kept fixed at $m_h$
and $m_h/4$, respectively.
We note that in Fig.~\ref{fig:csrsvar}, the pure QED contributions are large.
This is due to the fact that we consider leading order QCD running of Yukawa coupling 
which gives larger Born contribution compared to pure QCD.   
In order to understand this in more detail, we study the impact of different contributions
to the cross sections resulting from QCD, QED and mixed QCD$\times$QED
at various orders in perturbation theory which we have tabulated 
in Table~\ref{tab:cont}, 
for $\sqrt{S}=14$ TeV and for the scale choice $\mu_R=\mu_F=m_h$.  
{
The $\Delta^{i,j}$ indicates sole $i$-th order QCD and $j$-th order QED corrections
to the total contribution. For example, 
NNLO$_{11}$ means $\Delta^{0,0} + \Delta^{1,0} + \Delta^{0,1} + \Delta^{1,1}$}.\\
\begin{table}[ht!]
\begin{tabular}{ l | c | c | c | c | c | c }
 \hline\hline\noalign{\smallskip}
     & $\Delta^{0,0}$ & $\Delta^{1,0}$ & $\Delta^{0,1}$ & $\Delta^{2,0}$ & $\Delta^{1,1}$ & $\Delta^{0,2}$ \\  
 \noalign{\smallskip}\hline\noalign{\smallskip} 
 LO$_{00}$   & 1.0181 &  &  & & & \\ 
 NLO$_{10}$  & 1.1362 & -0.1810 & & & & \\ 
 NLO$_{01}$  & 1.2219 &  &  0.0030 & & & \\
 NNLO$_{20}$  & 1.1433 & -0.1683 & & -0.1935 & & \\ 
 NNLO$_{11}$  & 1.1542 & -0.1699 & 0.0029 &  & -0.0005 & \\ 
 NNLO$_{02}$  & 1.2422 & & 0.0031 &  &  & -4 $10^{-6}$ \\
 \noalign{\smallskip} \hline\hline
\end{tabular}
\caption{Individual contributions in (pb) to various perturbative orders at $\sqrt{S}$=14 TeV.}
\label{tab:cont}
\end{table}
In Table~\ref{tab:cont2}, a similar study has been performed 
for $\sqrt{S}=13$ TeV and the scales $\mu_R=m_h\,,~\mu_F=m_h/4$.  
\begin{table}[ht!]
\begin{tabular}{ l | c | c | c | c | c | c }
 \hline\hline\noalign{\smallskip}
    & $\Delta^{0,0}$ & $\Delta^{1,0}$ & $\Delta^{0,1}$ & $\Delta^{2,0}$ & $\Delta^{1,1}$ & $\Delta^{0,2}$ \\  
 \noalign{\smallskip}\hline\noalign{\smallskip} 
 LO$_{00}$   & 0.3911 &  &  & & & \\ 
 NLO$_{10}$  & 0.4588 & 0.1557 & & & & \\ 
 NLO$_{01}$  & 0.4935 &  & 0.0003 & & & \\
 NNLO$_{20}$  & 0.4726 & 0.1614 & & 0.0220 & & \\ 
 NNLO$_{11}$  & 0.4771 & 0.1630 & 0.0003 &  & 1.5 $10^{-4}$ & \\ 
 NNLO$_{02}$  & 0.5135 & & 0.0003 &  &  & 6 $10^{-6}$ \\
 \noalign{\smallskip} \hline\hline
\end{tabular}
\caption{Individual contributions in (pb) to various perturbative orders at $\sqrt{S}$=13 TeV.}
\label{tab:cont2}
\end{table}

\begin{figure}[ht!]
\includegraphics[width=0.45\textwidth]{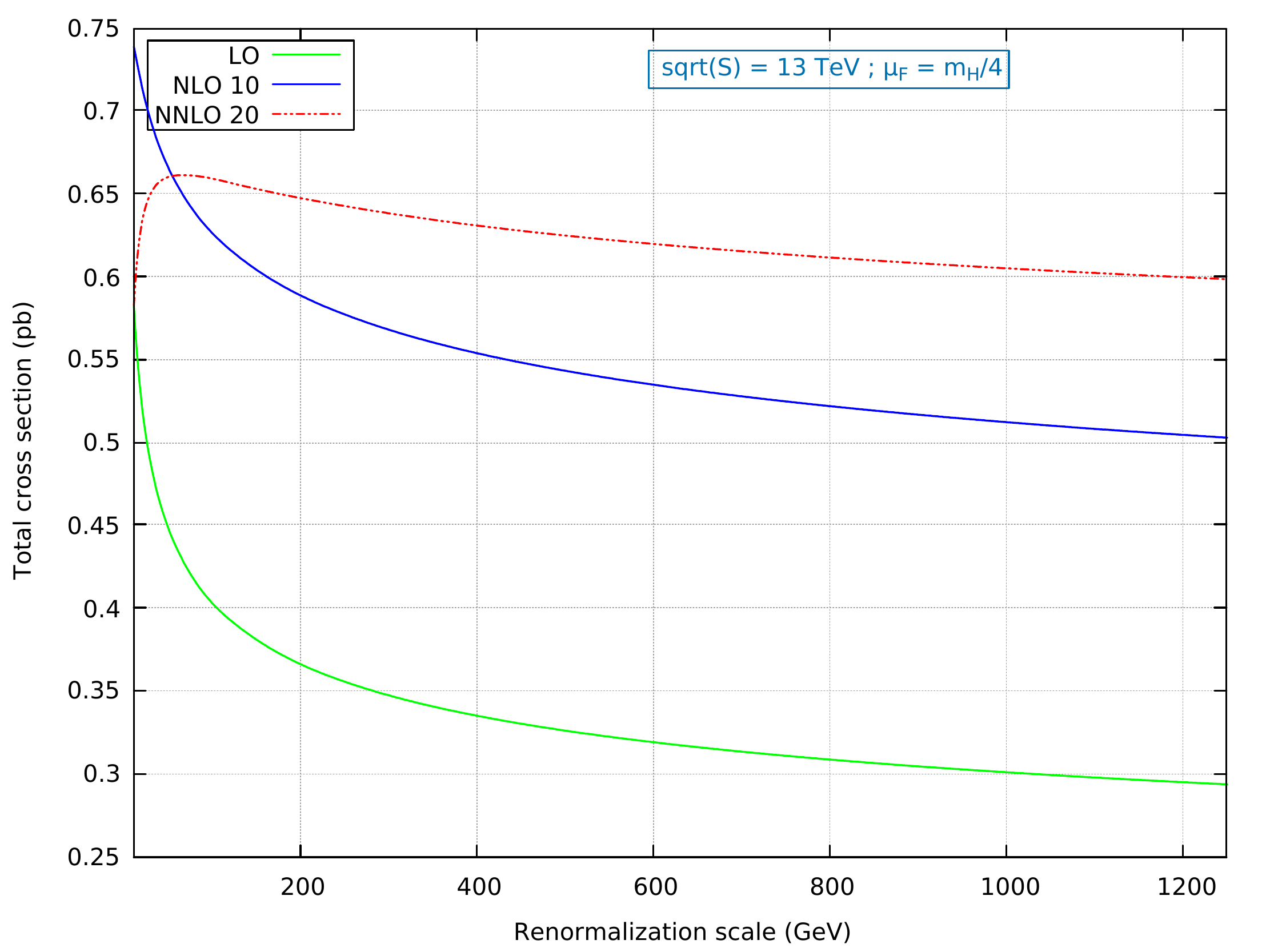}
\caption{\sf The renormalization scale variation of the total cross section at various perturbative orders in QCD.}
\label{fig:csmuRvar}
\end{figure}
Fixed order predictions depend on the renormalization ($\mu_R$) and factorization ($\mu_F$) scales.  The uncertainty resulting from the
choice of the scales quantify the missing higher order contributions.  Hence, we have studied their dependence by varying them independently around 
a central scale.  
{
Fig.~\ref{fig:csmuRvar} shows the dependence of the cross section on the renormalization scale ($\mu_R$) for the fixed choice of
the factorization scale $\mu_F=m_h/4$. It clearly demonstrates the importance of 
higher order corrections as 
the $\mu_R$ variation is much more stable at NNLO$_{20}$ compared to the lower orders.}
\begin{figure}[ht!]
\includegraphics[width=0.45\textwidth]{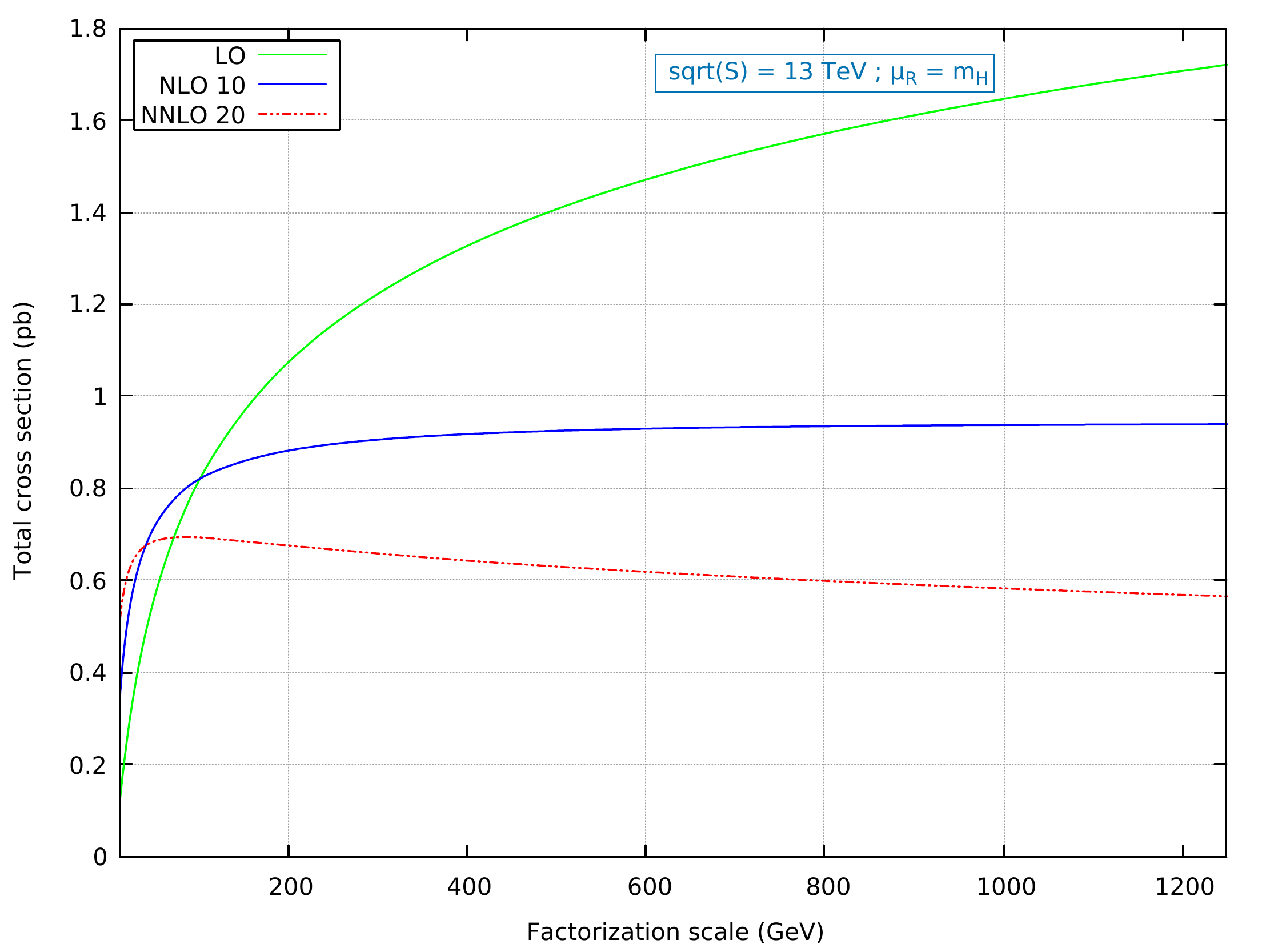}
\caption{\sf The factorization scale variation of the total cross section at various perturbative orders in QCD.}
\label{fig:csmuFvar}
\end{figure}
{
In Fig.~\ref{fig:csmuFvar}, we present the dependence on the factorization scale ($\mu_F$) keeping 
the renormalization scale ($\mu_R$) fixed at $m_h$.  Similar to the $\mu_R$ variation, $\mu_F$ variation
improves after adding higher order corrections.}
To illustrate their dependence when both the scales are changed simultaneously, 
we present the cross section by performing 7-point scale variation and the results are listed in 
Table~\ref{tab:7p1}. 
We have used {\small \sf NNPDF31\_nnlo\_as\_0118\_luxqed} for this study.
\begin{table}[ht!]
\begin{tabular}{ l | c | c | c | c | c | c | c }
 \hline\hline\noalign{\smallskip}
  ($\frac{\mu_R}{m_h},\frac{\mu_F}{m_h}$)  & (2,$\frac{1}{2}$) & (2,$\frac{1}{4}$) & (1,$\frac{1}{2}$) & (1,$\frac{1}{4}$) & (1,$\frac{1}{8}$) & ($\frac{1}{2}$,$\frac{1}{4}$) & ($\frac{1}{2}$,$\frac{1}{8}$) \\  
 \noalign{\smallskip}\hline\noalign{\smallskip} 
 NNLO$_{20}$ (pb) & 0.707 & 0.643 & 0.690 & 0.656 & 0.562 & 0.661 & 0.606 \\ 
 NNLO$_{11}$ (pb) & 0.759 & 0.602 & 0.780 & 0.641 & 0.445 & 0.682 & 0.498 \\ 
 NNLO$_{02}$ (pb) & 0.728 & 0.465 & 0.804 & 0.514 & 0.250 & 0.574 & 0.279 \\ 
 \noalign{\smallskip} \hline\hline
\end{tabular}
\caption{7-point scale variation at $\sqrt{S}$=13 TeV.}
\label{tab:7p1}
\end{table}

The perturbative predictions also depend on the choice of PDFs and PHDFs.
There are several groups which fit them and are widely used in the literature for the phenomenological studies.
In order to estimate the uncertainty resulting from the choice of PDFs and PHDFs,  in Table~\ref{tab:pdf1}, we present 
the NNLO results from various PDF sets, for $\sqrt{S}$=14 TeV and $\mu_R = \mu_F = m_h$. 
\begin{table}[ht]
\begin{tabular}{ c | c | c | c | c }
 \hline\hline\noalign{\smallskip}
              & MRST & NNPDF & CT14 & PDF4LHC  \\
 \noalign{\smallskip}\hline\noalign{\smallskip}              
 NNLO$_{20}$ (pb) & 0.7805 & 0.7816 & 0.7574 & 0.8546 \\ 
 NNLO$_{11}$ (pb) & 0.9691 & 0.9867 & 0.9644 & 1.0625 \\   
 NNLO$_{02}$ (pb) & 1.2020 & 1.2453 & 1.2288 & 1.3123 \\      
 \noalign{\smallskip} \hline\hline
\end{tabular}
\caption{Result using different PDFs at $\sqrt{S}$=14 TeV.}
\label{tab:pdf1}
\end{table}
In Table~\ref{tab:pdf2}, we repeat the study 
for $\sqrt{S}$=13 TeV and $\mu_R = m_h$ and $\mu_F = m_h/4$.
\begin{table}[ht!]
\begin{tabular}{ c | c | c | c | c }
 \hline\hline\noalign{\smallskip}
              & MRST & NNPDF & CT14 & PDF4LHC  \\
 \noalign{\smallskip}\hline\noalign{\smallskip}              
 NNLO$_{20}$ (pb) & 0.6610 & 0.6561 & 0.6398 & 0.7178 \\
 NNLO$_{11}$ (pb) & 0.6451 & 0.6406 & 0.6259 & 0.6996 \\   
 NNLO$_{02}$ (pb) & 0.5252 & 0.5139 & 0.5030 & 0.5605 \\      
 \noalign{\smallskip} \hline\hline
\end{tabular}
\caption{Result using different PDFs at $\sqrt{S}$=13 TeV.}
\label{tab:pdf2}
\end{table}
We have also studied the uncertainties resulting from the choice of PDF set 
\cite{Buckley:2014ana}.
Using NNPDF31, in Fig~\ref{fig:pdf}, we plot the variation of the cross section 
with respect to different choices of PDF and PHDF templates keeping the central set 
as the reference.  The thick line is obtained using the central set. 
The shaded region resulting from other sets quantifies the uncertainty.
\begin{figure}[ht!]
\includegraphics[width=0.4\textwidth]{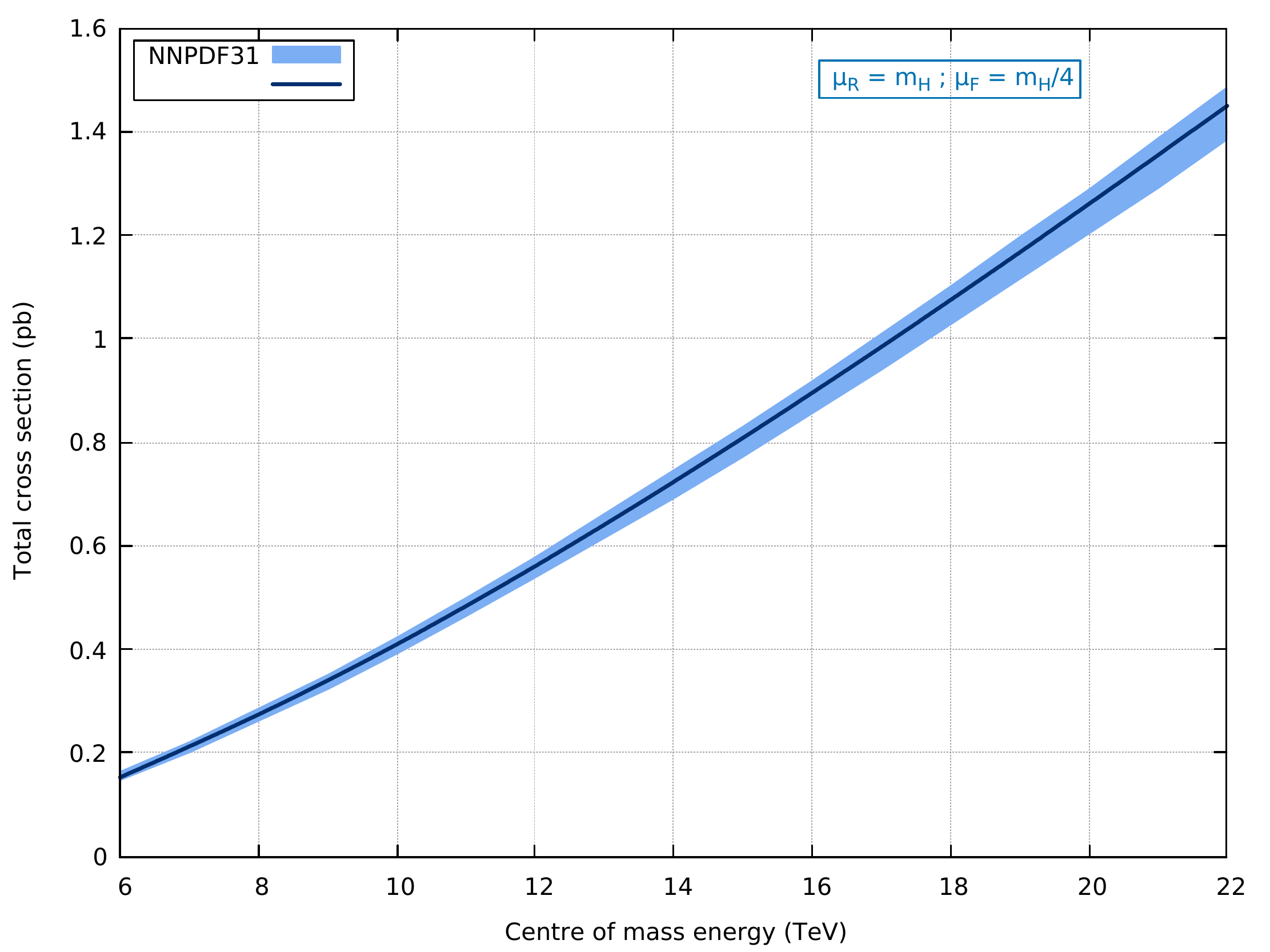}
\caption{\sf PDF uncertainties.}
\label{fig:pdf}
\end{figure}

\section{Discussion and Conclusion}
\label{sec:conclusion}
\noindent
Precision studies is one of the prime areas at the LHC.  Measuring the parameters of
the SM to unprecedented accuracy can help us to improve our understanding of the
dynamics that governs the particle interactions at high energies.  This is possible
only if the accuracy of theoretical predictions is comparable to that of the measurements.  
Thanks to the on-going efforts from experimentalists and theorists, there are stringent constraints
on various physics scenarios in the pursuit of searching for the physics beyond the SM.  
The efforts to compute the observables that are related to top quarks and Higgs bosons have been going on 
for a while as these observables are sensitive to high scale physics.  Since the dominant
contributions to these processes are known to unprecedented accuracy, inclusion of 
sub-dominant contributions along with radiative corrections is essential for any consistent
study.  In this context, the present
article explores the possibility of including EW corrections to Higgs boson production
in bottom quark annihilation which is sub-dominant.  Note that, this is known
to third order in QCD \cite{Duhr:2019kwi}.   While this is a sub-dominant process at the LHC, in certain 
BSM contexts, the rates are significantly appreciable leading to interesting phenomenological
studies.  Since, the computation of full EW corrections is more involved, as a first step
towards this, we compute all the QED corrections, in particular, to the inclusive Higgs boson production
in bottom quark annihilation up to second order in QED coupling constant $a_e$, taking
into account the non-factorizable or mixed QCD$\times$QED effects through $a_s a_e$ corrections.
The computation involves dealing with QED soft and collinear singularities resulting from photons
and the massless partons along with the corresponding QCD ones.  Understanding the 
structure of these QED IR singularities in the presence of QCD ones, is a challenging task.
We have systematically investigated both QCD and QED IR singularities up to second order in
their couplings taking into account the interference effects.  We use Sudakov K+G equation 
to understand the IR structure in terms of cusp, colliner and soft anomalous dimensions.   
We demonstrate that the IR singularities from QCD, QED and QCD$\times$QED interactions factorize
both at the FF, as well as at the cross section level.  While the IR singularities factorize as a whole, 
the IR singularities from QCD do not factorize from that of QED leading to mixed/non-factorizable 
QCD$\times$QED IR singularities.  In addition, by computing the real emission processes in the limit
when the photons/gluons become soft, we have studied the structure of soft distribution function.
While the later demonstrates the universal structure analogous to QCD one,
we find that it contains soft terms from mixed QCD$\times$QED that do not factorize either as a product of
those from QCD and QED separately.  Using the universal IR structure of the observable, we have determined
the mass anomalous dimension of the bottom quark and hence the renormalization constant for the
bottom Yukawa.  We also discussed the relation between the results 
from pure QED and pure QCD as well as between QCD$\times$QED and QCD through Abelianization.  
We have determined a set of
rules that relate them and they are found to be consistent with those observed in the context of DY 
\cite{deFlorian:2018wcj}.
Having obtained the complete NNLO results from QED and QCD$\times$QED, we have systematically
included them in the NNLO QCD study to understand their impact at the LHC energy.  We find that
the corrections are mild as expected.  However, we show that the higher order corrections
from QED and QCD$\times$QED improve the reliability of the predictions.  

\section*{Acknowledgment}
A.A.H, A.C, P.M and V.R  would like to acknowledge the support of the 
CNRS LIA (Laboratoire International Associé) THEP (Theoretical High Energy Physics) 
and the INFRE-HEPNET (IndoFrench Network on High Energy Physics) 
of CEFIPRA/IFCPAR (Indo-French Centre for the Promotion of Advanced Research).
We thank G. Ferrera and A. Vicini for useful discussions.

\appendix
\section{\texorpdfstring{$\mathcal{\overline G}_{I,ij}^{(k)}$}{giik}s of the soft distribution function} \label{ap:g}
\noindent
The constants $\mathcal{\overline G}_{I,ij}^{(k)}$ in the soft distribution function are given by,
\begin{align}
\mathcal{\overline G}_{I,10}^{(1)} &= C_F \Big( -3 \zeta_2 \Big) \,, \nonumber \\ 
\mathcal{\overline G}_{I,10}^{(2)} &= C_F \Big( \frac{7}{3}\zeta_3 \Big) \,, \nonumber \\
\mathcal{\overline G}_{I,10}^{(3)} &= C_F \Big( -\frac{3}{16}\zeta_2^2 \Big) \,, \nonumber \\
\mathcal{\overline G}_{I,01}^{(1)} &= e_b^2 \Big( -3 \zeta_2 \Big) \,, \nonumber \\
\mathcal{\overline G}_{I,01}^{(2)} &= e_b^2 \Big( \frac{7}{3}\zeta_3 \Big) \,, \nonumber \\
\mathcal{\overline G}_{I,01}^{(3)} &= e_b^2 \Big( -\frac{3}{16}\zeta_2^2 \Big) \,, \nonumber \\
\mathcal{\overline G}_{I,11}^{(1)} & = 0 \,,       \nonumber \\ 
\mathcal{\overline G}_{I,20}^{(1)} &=  C_F n_f T_F \Big (-\frac{656}{81} + \frac{140}{9}\zeta_2 + \frac{64}{3} \zeta_3 \Big)  
+ \nonumber \\ 
& \hspace{.5cm}   C_A C_F \Big( \frac{2428}{81} -\frac{469}{9} \zeta_2 + 4 \zeta_2^2 - \frac{176}{3} \zeta_3 \Big) \,, \nonumber \\
\mathcal{\overline G}_{I,02}^{(1)} &= e_b^2 \Big( N \sum_{q}e_{q}^2\Big) \Big (-\frac{656}{81} + \frac{140}{9}\zeta_2 + \frac{64}{3} \zeta_3 \Big) \,. 
\end{align}

\section{Form factors} \label{ap:ff}
We present the analytic expressions of the form factors and 
the finite partonic cross sections for all the partonic channels. The labeling is same as Fig.~\ref{fig:csrsvar}
\begin{widetext}
\newcommand{\f}{{\cal F}}
\noindent
The unrenormalized form factor ($\hat{F}_I$) can be written as follows in the perturbative expansion of unrenormalized strong coupling constant ($\hat{a}_s$)
and unrenormalized fine structure constant ($\hat{a}_e$)
\begin{align}
 \hat{F}_I &= 1 + \hat{a}_s \Big(\frac{Q^2}{\mu^2}\Big)^{\frac{\varepsilon}{2}} {\cal S}_{\varepsilon} \Big[ C_F \f_1^I \Big]
              + \hat{a}_e \Big(\frac{Q^2}{\mu^2}\Big)^{\frac{\varepsilon}{2}} {\cal S}_{\varepsilon} \Big[ e_I^2 \f_1^I \Big]
              + \hat{a}_s^2 \Big(\frac{Q^2}{\mu^2}\Big)^{\varepsilon} {\cal S}_{\varepsilon}^2 \Big[ C_F^2 \f_{2,0}^I + C_A C_F \f_{2,1}^I + C_F n_f T_F \f_{2,2}^I \Big]
\nonumber\\&             
              + \hat{a}_s \hat{a}_e  \Big(\frac{Q^2}{\mu^2}\Big)^{\varepsilon} {\cal S}_{\varepsilon}^2 \Big[ 2 C_F e_I^2 \f_{2,0}^I \Big]
              + \hat{a}_e^2 \Big(\frac{Q^2}{\mu^2}\Big)^{\varepsilon} {\cal S}_{\varepsilon}^2 \Big[ e_I^4 \f_{2,0}^I + e_I^2 \Big( N \sum_q e_q^2 \Big) \f_{2,2}^I \Big] \,.
\end{align}
$I=q,b$ denotes the Drell-Yan pair production and the Higgs boson production in bottom quark annihilation, respectively. 
The coefficients $\f_1^q, \f_{2,0}^q, \f_{2,1}^q$ and $\f_{2,2}^q$ are
\begin{align}
 \f_1^q &= - \frac{8}{\varepsilon^2} 
 + \frac{6}{\varepsilon} - 8 + \zeta_2 
 +\varepsilon \Big( 8 - \frac{3}{4}\zeta_2 - \frac{7}{3}\zeta_3 \Big) 
 +\varepsilon^2 \Big( -8 + \zeta_2 + \frac{47}{80}\zeta_2^2 + \frac{7}{4}\zeta_3 \Big) 
 +\varepsilon^3 \Big( 8 - \zeta_2 - \frac{141}{320}\zeta_2^2 - \frac{7}{3}\zeta_3  
\nonumber\\
&
                    + \frac{7}{24}\zeta_2 \zeta_3 - \frac{31}{20}\zeta_5 \Big) 
 + \varepsilon^4 \Big( -8 + \zeta_2 + \frac{47}{80}\zeta_2^2 + \frac{949}{4480}\zeta_2^3
 + \frac{7}{3}\zeta_3 -\frac{7}{32}\zeta_2\zeta_3 -\frac{49}{144}\zeta_3^2 + \frac{93}{80}\zeta_5 \Big) \,.
\\
\f_{2,0}^q &= \frac{32}{\varepsilon^4} - \frac{48}{\varepsilon^3}
            + \frac{1}{\varepsilon^2}\Big( 82 - 8 \zeta_2 \Big)
            + \frac{1}{\varepsilon} \Big( -\frac{221}{2} + \frac{128}{3}\zeta_3  \Big)
            + \frac{1151}{8} + \frac{17}{2}\zeta_2 - 13 \zeta_2^2 - 58 \zeta_3   
+ \varepsilon \Big(-\frac{5741}{32}-\frac{213}{8}\zeta_2 
\nonumber\\
& 
                  + \frac{171}{10}\zeta_2^2 +\frac{839}{6}\zeta_3 -\frac{56}{3}\zeta_2\zeta_3 + \frac{92}{5}\zeta_5 \Big)
+ \varepsilon^2 \Big(\frac{27911}{128} + \frac{1839}{32}\zeta_2 - \frac{3401}{80}\zeta_2^2 + \frac{223}{20}\zeta_2^3 - \frac{6989}{24}\zeta_3 
+ \frac{27}{2}\zeta_2\zeta_3 
\nonumber\\ &
+ \frac{652}{9}\zeta_3^2 - \frac{231}{10}\zeta_5 \Big) \,.
\\
\f_{2,1}^q &= \frac{1}{\varepsilon^3} \Big( \frac{44}{3} \Big)
+ \frac{1}{\varepsilon^2} \Big( -\frac{332}{9} + 4 \zeta_2 \Big)
+ \frac{1}{\varepsilon} \Big( \frac{4129}{54} + \frac{11}{3}\zeta_2 - 26\zeta_3 \Big)
-\frac{89173}{648} - \frac{119}{9}\zeta_2 + \frac{44}{5}\zeta_2^2 + \frac{467}{9}\zeta_3  
\nonumber\\
&
+\varepsilon \Big(\frac{1775893}{7776} + \frac{6505}{216}\zeta_2 - \frac{1891}{120}\zeta_2^2 -\frac{3293}{27}\zeta_3 + \frac{89}{6}\zeta_2\zeta_3 -\frac{51}{2}\zeta_5  \Big) 
+ \varepsilon^2 \Big(-\frac{33912061}{93312} - \frac{146197}{2592}\zeta_2 + \frac{2639}{72}\zeta_2^2 
\nonumber\\
&- \frac{809}{280}\zeta_2^3 + \frac{159949}{648}\zeta_3 
- \frac{397}{36}\zeta_2\zeta_3 -\frac{569}{12}\zeta_3^2 + \frac{3491}{60}\zeta_5  \Big) \,.
\\
\f_{2,2}^q &= \frac{1}{\varepsilon^3} \Big(-\frac{16}{3} \Big)
+ \frac{1}{\varepsilon^2} \Big( \frac{112}{9} \Big)
+ \frac{1}{\varepsilon} \Big( -\frac{706}{27} -\frac{4}{3} \zeta_2 \Big)
+ \frac{7541}{162} + \frac{28}{9}\zeta_2 -\frac{52}{9}\zeta_3 
+\varepsilon \Big( -\frac{150125}{1944} - \frac{353}{54}\zeta_2 + \frac{41}{30}\zeta_2^2 + \frac{364}{27}\zeta_3 \Big)
\nonumber \\ &
+ \varepsilon^2 \Big( \frac{2877653}{23328} + \frac{7541}{648}\zeta_2 -\frac{287}{90}\zeta_2^2 -\frac{4589}{162}\zeta_3 
- \frac{13}{9}\zeta_2\zeta_3 - \frac{121}{15}\zeta_5  \Big) \,.
\end{align}
The coefficients $\f_1^b, \f_{2,0}^b, \f_{2,1}^b$ and $\f_{2,2}^b$ are
\begin{align}
 \f_1^b &= - \frac{8}{\varepsilon^2} - 2 + \zeta_2 
+ \varepsilon \Big( 2 - \frac{7}{3}\zeta_3 \Big) 
+ \varepsilon^2 \Big( -2 + \frac{1}{4}\zeta_2 + \frac{47}{80} \zeta_2^2 \Big)
+  \varepsilon^3 \Big( 2 - \frac{1}{4}\zeta_2 - \frac{7}{12} \zeta_3 + \frac{7}{24}\zeta_2 \zeta_3 - \frac{31}{20} \zeta_5  \Big)
\nonumber \\ &
+  \varepsilon^4 \Big( -2 + \frac{1}{4}\zeta_2 + \frac{47}{320} \zeta_2^2 + \frac{949}{4480} \zeta_2^3 + \frac{7}{12} \zeta_3 - \frac{49}{144} \zeta_3^2  \Big) \,.
\\
\f_{2,0}^b &= \frac{32}{\varepsilon^4}
+ \frac{1}{\varepsilon^2} \Big( 16 - 8 \zeta_2 \Big) 
+ \frac{1}{\varepsilon} \Big( -16-12\zeta_2 + \frac{128}{3}\zeta_3 \Big)
+ 22 + 12\zeta_2 - 13 \zeta_2^2 - 30 \zeta_3 
+ \varepsilon \Big( -32 - 18 \zeta_2 + \frac{48}{5} \zeta_2^2 
\nonumber\\&
+ \frac{202}{3} \zeta_3 
- \frac{56}{3} \zeta_2 \zeta_3 + \frac{92}{5} \zeta_5 \Big)
+ \varepsilon^2 \Big( 48 + \frac{53}{2} \zeta_2 - \frac{213}{10} \zeta_2^2 + \frac{223}{20} \zeta_2^3
- \frac{436}{3} \zeta_3 + \frac{1}{2} \zeta_2 \zeta_3 + \frac{652}{9} \zeta_3^2 - \frac{63}{2} \zeta_5 \Big) \,.
\\
\f_{2,1}^b &= \frac{1}{\varepsilon^3}\Big( \frac{44}{3} \Big)
+ \frac{1}{\varepsilon^2} \Big( -\frac{134}{9} + 4 \zeta_2 \Big)
+ \frac{1}{\varepsilon} \Big( \frac{440}{27} + \frac{11}{3}\zeta_2 - 26 \zeta_3 \Big)
 - \frac{1655}{81} -\frac{103}{18}\zeta_2 + \frac{44}{5}\zeta_2^2 + \frac{305}{9}\zeta_3 
\nonumber \\ &
+\varepsilon \Big( \frac{6353}{243} + \frac{245}{27} \zeta_2 - \frac{1171}{120} \zeta_2^2 
- \frac{2923}{54} \zeta_3 
+ \frac{89}{6} \zeta_2 \zeta_3 - \frac{51}{2} \zeta_5 \Big)
+ \varepsilon^2 \Big( -\frac{49885}{1458} - \frac{4733}{324} \zeta_2 + \frac{11819}{720} \zeta_2^2 
-\frac{809}{280} \zeta_2^3 
\nonumber\\&
+ \frac{7667}{81} \zeta_3 
- \frac{127}{36} \zeta_2 \zeta_3 
-\frac{569}{12} \zeta_3^3 + \frac{2411}{60} \zeta_5 \Big) \,. 
\\
\f_{2,2}^b &= \frac{1}{\varepsilon^3} \Big( -\frac{16}{3} \Big)
+ \frac{1}{\varepsilon^2} \Big( \frac{40}{9} \Big)
+ \frac{1}{\varepsilon} \Big( -\frac{184}{27} -\frac{4}{3} \zeta_2 \Big)
 + \frac{832}{81} + \frac{10}{9}\zeta_2 -\frac{52}{9}\zeta_3 
+\varepsilon \Big( -\frac{3748}{243} - \frac{46}{27}\zeta_2 + \frac{41}{30} \zeta_2^2 + \frac{130}{27} \zeta_3 \Big)
\nonumber \\ &
+ \varepsilon^2 \Big( \frac{16870}{729} + \frac{208}{81} \zeta_2 - \frac{41}{36} \zeta2^2 
-\frac{598}{81} \zeta_3 - \frac{13}{9} \zeta_2 \zeta_3 - \frac{121}{15} \zeta_5 \Big) \,.
\end{align}

\section{\texorpdfstring{$\Delta_{cd}^{(i,j)}$}{delcd} for bottom quark annihilation from QCD, QED and QCD\texorpdfstring{$\times$}{tt}QED up to NNLO}
\label{ap:cs}
\noindent
In the following, we present finite partonic cross sections $\Delta_{cd}^{(i,j)}$ as defined in Eq.~\ref{eq:delexp}, up to NNLO level in the strong and 
electro-magnetic coupling constants.  In QCD, $\Delta_{cd}^{i,0}$ for 
bottom quark annihilation is already known \cite{Harlander:2003ai,Majhi:2010zg}.  In the following,  $\Delta_{cd}^{i,0}, i=1,2$ is
in $SU(N)$ gauge theory, while $\Delta_{cd}^{0,j},j=1,2$ is in $U(1)$ gauge theory. 
\begin{align}
\Delta_{b\bar{b}}^{(0,0)} &= \delta (1-z) \,.
% % % % % % % % 
\\
\Delta_{b\bar{b}}^{(1,0)} &= C_F \Big\{ 
\delta (1-z) \Big( -4 + 8 \zeta_2 \Big)
+ 16 \XD_1
+ 4 (1-z)
-8 (1+z) \log (1-z)
-\frac{4 \big(
        1+z^2\big)}{(1-z)} \log (z)
\Big\}    \,.
\\
\Delta_{bg}^{(1,0)} &=  
-\frac{1}{2} (-1+z) (-3+7 z)
+2 \big(
        1-2 z+2 z^2\big) \log (1-z)
+\big(
        -1+2 z-2 z^2\big) \log (z) \,.
% 
% % % % % % % % % % % % % % % % % % % % % % % % % % 
\\
\Delta_{b\bar{b}}^{(2,0)} &= 
C_F^2 \Big\{
\delta(1-z) \Big(
16
+\frac{8}{5} \zeta_2^2
-60 \zeta_3
\Big)
+256 \XD_0 \zeta_3
+\XD_1 \big(
        -64
        -128 \zeta_2
\big)
+ 128 \XD_3
% 
%%%%%%%%%%
-4 \big(
        -26+11 z+13 z^2\big)
\nonumber\\&
        +
\frac{8}{1-z} \big(
        -7-10 z+11 z^2\big) \log (1-z) \log (z)
-\frac{4}{1-z} \big(
        23+39 z^2\big) \log ^2(1-z) \log (z)
+\frac{2}{1-z} \big(
        7+30 z-34 z^2
\nonumber\\&        
        +12 z^3\big) \log ^2(z)
+\frac{16}{1-z} \big(
        2+5 z^2\big) \log (1-z) \log ^2(z)
-\frac{2}{3 (1-z)} \big(
        1+15 z^2+4 z^3\big) \log ^3(z)
+\frac{8}{1-z} \big(
        -16+13 z
\nonumber\\&        
        -6 z^2+6 z^3\big) \text{Li}_2(1-z)
-\frac{8}{1-z} \big(
        -7+9 z^2\big) \log (1-z) \text{Li}_2(1-z)
-\frac{16}{1-z} \big(
        3+z^2+2 z^3\big) \log (z) \text{Li}_2(1-z)
\nonumber\\&        
+\frac{48}{1-z} \big(
        -1+2 z^2\big) \text{Li}_3(1-z)
-\frac{8}{1-z} \big(
        9+9 z^2+8 z^3\big) \text{S}_{1,2}(1-z)
-8 (-11+10 z) \zeta_2
-\frac{16}{1-z} \big(
        -2-7 z^2
\nonumber\\&        
        +z^3\big) \log (z) \zeta_2
-128 (1+z) \zeta_3
+12 (-4+9 z) \log (1-z)
+64 (1+z) \zeta_2 \log (1-z)
-32 (1-z) \log ^2(1-z)
\nonumber\\&
-64 (1+z) \log ^3(1-z)
+\frac{4}{1-z} \big(
        16-z+z^2\big) \log (z)
-48 z^2 \zeta_2 \log (1+z)
+16 (-1+2 z) \log (z) \log (1+z)
\nonumber\\&
+40 z^2 \log ^2(z) \log (1+z)
-48 z^2 \log (z) \log ^2(1+z)
+16 (-1+2 z) \text{Li}
_2(-z)
+48 z^2 \log (z) \text{Li}_2(-z)
\nonumber\\&
-96 z^2 \log (1+z) \text{Li}_2(-z)
-16 z^2 \text{Li}_3(-z)
-96 z^2 \text{S}_{1,2}(-z)
\Big\}
% % % % 
+ C_A C_F \Big\{
\delta(1-z) \Big(
        \frac{166}{9}
        +\frac{232}{9} \zeta_2
        -\frac{12}{5} \zeta_2^2
        -8 \zeta_3
\Big)
\nonumber\\&
+\XD_0 \Big(
        -\frac{1616}{27}
        +\frac{176 \zeta_2}{3}
        +56 \zeta_3
\Big)
+\XD_1 \Big(
        \frac{1072}{9}
        -32 \zeta_2
\Big)
-\frac{176}{3} \XD_2
+ \frac{2}{27} \big(
        -595+944 z+351 z^2\big)
\nonumber\\&        
-\frac{4}{3 (1-z)} \big(
        61-31 z+40 z^2\big) \log (z)
+\frac{32}{3 (1-z)} \big(
        7+4 z^2\big) \log (1-z) \log (z)
-\frac{1}{3 (1-z)} \big(
        61+48 z-13 z^2
\nonumber\\&        
        +36 z^3\big) \log ^2(z)
+\frac{8}{(1-z)} \big(
        1+z^2\big) \log (1-z) \log ^2(z)
+\frac{2}{3 (1-z)} \big(
        -3-7 z^2+2 z^3\big) \log ^3(z)
\nonumber\\&        
-\frac{4}{3 (1-z)} \big(
        -29+27 z-27 z^2+18 z^3\big) \text{Li}_2(1-z)
+\frac{8}{(1-z)} \big(
        1+z^2\big) \log (1-z) \text{Li}_2(1-z)
\nonumber\\&        
+\frac{8}{(1-z)} \big(
        3+2 z^3\big) \log (z) \text{Li}_2(1-z)
-\frac{28}{(1-z)} \big(
        1+z^2\big) \text{Li}_3(1-z)
+\frac{8}{(1-z)} (1+z) \big(
        5-5 z+4 z^2\big) \text{S}_{1,2}(1-z)
\nonumber\\&
-\frac{4}{3} (22+25 z) \zeta_2
-28 (1+z) \zeta_3
-\frac{4}{9} (-40+299 z) \log (1-z)
+16 (1+z) \zeta_2 \log (1-z)
+\frac{88}{3} (1+z) \log ^2(1-z)
\nonumber\\&
+\frac{8}{(1-z)} (1+z) \big(
        1-z+z^2\big) \zeta_2 \log (z)
+24 z^2 \zeta_2 \log (1+z)
-8 (-1+2 z) \log (z) \log (1+z)
\nonumber\\&
-20 z^2 \log ^2(z) \log (1+z)
+24 z^2 \log (z) \log ^2(1+z)
-8 (-1+2 z) \text{Li}
_2(-z)
-24 z^2 \log (z) \text{Li}_2(-z)
\nonumber\\&
+48 z^2 \log (1+z) \text{Li}_2(-z)
+8 z^2 \text{Li}_3(-z)
+48 z^2 \text{S}_{1,2}(-z)
\Big\}
% % % % 
+ C_F n_f T_F \Big\{
\delta(1-z) \Big(
\frac{16}{9}
-\frac{80}{9} \zeta_2
+16 \zeta_3
\Big)
\nonumber\\&
+\XD_0 \Big(
        \frac{448}{27}
        -\frac{64}{3} \zeta_2
\Big)
-\frac{320}{9}  \XD_1
+\frac{64}{3} \XD_2
% 
%%%%%%%%
-\frac{8}{27} (1+55 z)
+\frac{8}{3 (1-z)} \big(
        7-4 z+7 z^2\big) \log (z)
\nonumber\\&        
-\frac{64}{3 (1-z)} \big(
        1+z^2\big) \log (1-z) \log (z)
+\frac{4}{3 (1-z)} \big(
        5+7 z^2\big) \log ^2(z)
+\frac{32}{3} (1+z) \zeta_2
+\frac{64}{9} (1+4 z) \log (1-z)
\nonumber\\&
-\frac{32}{3} (1+z) \log ^2(1-z)
-\frac{16}{3 (1-z)} \text{Li}_2(1-z)
\Big\}
%%%%%%%%%%%%%%% 
% 
+C_F T_F \Big\{
\frac{2}{27 z} (-1+z) \big(
        208-635 z+487 z^2\big)
\nonumber\\&        
-\frac{16}{9 z} (-1+z) \big(
        4-53 z+22 z^2\big) \log (1-z)
-\frac{16}{3 z} (-1+z) \big(
        4+7 z+4 z^2\big) \log ^2(1-z)
\nonumber\\&
+16 \big(
        -1+4 z+4 z^2\big) \log (1-z) \log (z)
+\frac{8}{3 z} \big(
        16-3 z+21 z^2+8 z^3\big) \text{Li}_2(1-z)
\nonumber\\&        
+64 (1+z) \log (1-z) \text{Li}_2(1-z)
-\frac{16}{3 z} \big(
        4+3 z-3 z^2-3 z^3\big) \zeta_2
+\frac{4}{9} \big(
        87-252 z+38 z^2\big) \log (z)
\nonumber\\&        
-32 (1+z) \zeta_2 \log (z)
+32 (1+z) \log ^2(1-z) \log (z)
-2 \big(
        1+5 z+12 z^2\big) \log ^2(z)
-32 (1+z) \log (1-z) \log ^2(z)
\nonumber\\&
+\frac{20}{3} (1+z) \log ^3(z)
-\frac{32}{3} z^2 \log (z) \log (1+z)
-16 (1+z) \log (z) \text{Li}_2(1-z)
\nonumber\\&
-\frac{32}{3} z^2 \text{Li}_2(-z)
-64 (1+z) \text{Li}_3(1-z)
+32 (1+z) \text{S}_{1,2}(1-z)
\Big\} \,.
\\
%%%%%%%%%%%%%%%%%%%%%%%% % % % % % % % % % % 
\Delta_{bb}^{(2,0)} &= 
C_F^2 \Big\{
        -2 (-1+z) (-57+13 z)
        +
        \frac{16}{1+z} \big(
                1+z^2\big) \log (1-z) \log ^2(z) 
        -\frac{4}{3 (1+z)} \big(
                3+7 z^2+2 z^3\big) \log ^3(z) 
\nonumber\\
&
        +\frac{4}{1+z} \big(
                9+19 z^2\big) \log ^2(z) \log (1+z)
        -\frac{8}{1+z} \big(
                -1+5 z^2\big) \log (z) \log ^2(1+z)
        -8 \big(
                -7-5 z+3 z^2\big) \text{Li}_2(1-z)
\nonumber\\
&
        -\frac{16}{1+z} \big(
                -3-2 z^2+z^3\big) \log (z) \text{Li}_2(1-z)
        -\frac{64}{1+z} \big(
                1+z^2\big) \log (1-z) \text{Li}_2(-z)
        +\frac{8 \big( 5+11 z^2\big)}{1+z}  \log (z) \text{Li}_2(-z)
\nonumber\\
&
        -\frac{16}{1+z} \big(
                -1+5 z^2\big) \log (1+z) \text{Li}_2(-z)
        +\frac{8}{1+z} \big(
                -7-z-8 z^2+2 z^3\big) \text{Li}_3(1-z)
        -\frac{8}{1+z} \big(
                1+3 z^2\big) \text{Li}_3(-z)
\nonumber\\
&
        +\frac{64}{1+z} \big(
                1+z^2\big) \text{Li}_3\Big(
                \frac{1-z}{1+z}\Big)
        -\frac{64}{1+z} \big(
                1+z^2\big) \text{Li}_3\Big(
                -\frac{1-z}{1+z}\Big)
        -\frac{8}{1+z} \big(
                -9+z-4 z^2+2 z^3\big) \text{S}_{1,2}(1-z)
\nonumber\\
&
        -\frac{16}{1+z} \big(
                -1+5 z^2\big) \text{S}_{1,2}(-z)
        + 32 (1+z) \log (1-z) \log (z)
        -\frac{64}{1+z} \big(
                        1+z^2\big) \log (1-z) \log (z) \log (1+z)                 
\nonumber\\
&      
        +8 \zeta_2
        -\frac{32}{1+z} \big(
                1+z^2\big) \log (1-z) \zeta_2
        -\frac{8}{1+z} \big(
                -1+5 z^2\big) \log (1+z) \zeta_2
        -\frac{8}{1+z} \big(
                1+z^2\big) \zeta_3
        +64 (1-z) \log (1-z)
\nonumber\\
&
        +4 (-8+5 z) \log (z)
        +\frac{16}{1+z} \big(
                1+2 z^2\big) \zeta_2 \log (z)
        -4 \big(
                1+2 z+3 z^2\big) \log ^2(z)
        +16 \log (z) \log (1+z)
        +16 \text{Li}_2(-z)
\Big\}
\nonumber\\
&
+ C_A C_F \Big\{
        57
        -70 z
        +13 z^2
        -
        \frac{8}{1+z} \big(
                1+z^2\big) \log (1-z) \log ^2(z) 
        +\frac{2}{3 (1+z)} \big(
                3+7 z^2+2 z^3\big) \log ^3(z)
\nonumber\\
&
        +\frac{32}{1+z} \big(
                1+z^2\big) \log (1-z) \log (z) \log (1+z)
        -\frac{2}{1+z} \big(
                9+19 z^2\big) \log ^2(z) \log (1+z)
\nonumber\\
&
        +\frac{4}{1+z} \big(
                -1+5 z^2\big) \log (z) \log ^2(1+z)
        +4 \big(
                -7-5 z+3 z^2\big) \text{Li}_2(1-z)
        -\frac{8}{1+z} \big(
                3+2 z^2-z^3\big) \log (z) \text{Li}_2(1-z)
\nonumber\\
&
        +\frac{32}{1+z} \big(
                1+z^2\big) \log (1-z) \text{Li}_2(-z)
        -\frac{4}{1+z} \big(
                5+11 z^2\big) \log (z) \text{Li}_2(-z)
        +\frac{8}{1+z} \big(
                -1+5 z^2\big) \log (1+z) \text{Li}_2(-z)
\nonumber\\
&
        +\frac{4}{1+z} \big(
                7+z+8 z^2-2 z^3\big) \text{Li}_3(1-z)
        +\frac{4}{1+z} \big(
                1+3 z^2\big) \text{Li}_3(-z)
        -\frac{32}{1+z} \big(
                1+z^2\big) \text{Li}_3\Big(
                \frac{1-z}{1+z}\Big)
\nonumber\\
&
        +\frac{32}{1+z} \big(
                1+z^2\big) \text{Li}_3\Big(
                \frac{-1+z}{1+z}\Big)
        +\frac{4}{1+z} \big(
                -9+z-4 z^2+2 z^3\big) \text{S}_{1,2}(1-z)
        +\frac{8}{1+z} \big(
                -1+5 z^2\big) \text{S}_{1,2}(-z)
        -4 \zeta_2
\nonumber\\
&
        +\frac{16}{1+z} \big(
                1+z^2\big) \log (1-z) \zeta_2
        +\frac{4}{1+z} \big(
                -1+5 z^2\big) \log (1+z) \zeta_2
        +\frac{4}{1+z} \big(
                1+z^2\big) \zeta_3
        +32 (-1+z) \log (1-z)
\nonumber\\
&
        -2 (-8+5 z) \log (z)
        -\frac{8}{1+z} \big(
                1+2 z^2\big) \zeta_2 \log (z)
        -16 (1+z) \log (1-z) \log (z)
        +2 \big(
                1+2 z+3 z^2\big) \log ^2(z)
\nonumber\\
&
        -8 \log (z) \log (1+z)
        -8 \text{Li}_2(-z)
\Big\}
%
% \nonumber\\&
%%%%%
+ C_F T_F \Big\{
        \frac{2}{27 z} (-1+z) \big(
                208-707 z+703 z^2\big)
        -\frac{16}{9 z} (-1+z) \big(
                4-53 z
\nonumber\\&                
                +22 z^2\big) \log (1-z)
        -\frac{16}{3 z} (-1+z) \big(
                4+7 z+4 z^2\big) \log ^2(1-z)
        +16 \big(
                -1+4 z+4 z^2\big) \log (1-z) \log (z)
\nonumber\\&                
        -\frac{2}{3} \big(
                3+15 z+40 z^2\big) \log ^2(z)
        +\frac{8}{3 z} \big(
                16-3 z+21 z^2+8 z^3\big) \text{Li}_2(1-z)
        +64 (1+z) \log (1-z) \text{Li}_2(1-z)
\nonumber\\&        
        +\frac{16}{3 z} (-1+z) \big(
                4+7 z+4 z^2\big) \zeta_2
        +\frac{4}{9} \big(
                93-264 z+20 z^2\big) \log (z)
        -32 (1+z) \zeta_2 \log (z)
\nonumber\\&        
        +32 (1+z) \log ^2(1-z) \log (z)        
        -32 (1+z) \log (1-z) \log ^2(z)
        +\frac{20}{3} (1+z) \log ^3(z)
\nonumber\\&        
        -16 (1+z) \log (z) \text{Li}_2(1-z)        
        -64 (1+z) \text{Li}_3(1-z)
        +32 (1+z) \text{S}_{1,2}(1-z)
\Big\} \,.
\\
% % % % % % % % % % 
% 
\Delta_{ub}^{(2,0)} &= 
C_F T_F \Big\{
        \frac{1}{27 z} (-1+z) \big(
                208-707 z+703 z^2\big) 
        -\frac{8}{9 z} (-1+z) \big(
                4-53 z+22 z^2\big) \log (1-z)
\nonumber\\
&
        -\frac{8}{3 z} (-1+z) \big(
                4+7 z+4 z^2\big) \log ^2(1-z)
        +8 \big(
                -1+4 z+4 z^2\big) \log (1-z) \log (z)
        -\frac{1}{3} \big(
                3+15 z+40 z^2\big) \log ^2(z)
\nonumber\\
&                
        +\frac{4}{3 z} \big(
                16-3 z+21 z^2+8 z^3\big) \text{Li}_2(1-z)
        +32 (1+z) \log (1-z) \text{Li}_2(1-z)
        +\frac{8}{3 z} (-1+z) \big(
                4+7 z+4 z^2\big) \zeta_2
\nonumber\\
&
        +\frac{2}{9} \big(
                93-264 z+20 z^2\big) \log (z)
        -16 (1+z) \zeta_2 \log (z)
        +16 (1+z) \log ^2(1-z) \log (z)
        -16 (1+z) \log (1-z) \log ^2(z)
\nonumber\\
&
        +\frac{10}{3} (1+z) \log ^3(z)
        -8 (1+z) \log (z) \text{Li}_2(1-z)
        -32 (1+z) \text{Li}_3(1-z)
        +16 (1+z) \text{S}_{1,2}(1-z)
\Big\}\,.
\\
% % % % % % % % % % % % % % % 
% 
\Delta_{u\bar{u}}^{(2,0)} &= 
C_F T_F \Big\{
        -\frac{16}{3} (-1+z) (-1+3 z)
        -\frac{16}{3} z^2 \zeta_2
        +\frac{8}{3} (1+z) (-1+3 z) \log (z)
        +\frac{8}{3} z^2 \log ^2(z)
\nonumber\\
&         
        -\frac{32}{3} z^2 \log (z) \log (1+z)       
        -\frac{32}{3} z^2 \text{Li}_2(-z)
\Big\}\,.
% 
%%%%%%%%%%%%%%%%%%%%%%%%%%
\\
% % 
\Delta_{bg}^{(2,0)} &= 
C_F \Big\{
\frac{1}{4} \big(
        -129+658 z-549 z^2\big)
+\big(
        64-197 z+136 z^2\big) \log (1-z)
-3 \big(
        11-32 z+23 z^2\big) \log ^2(1-z)
\nonumber\\&        
+\frac{35}{3} \big(
        1-2 z+2 z^2\big) \log ^3(1-z)
+4 \big(
        7-32 z+27 z^2\big) \log (1-z) \log (z)
-3 \big(
        7-14 z+22 z^2\big) \log ^2(1-z) \log (z)
\nonumber\\&        
+\frac{1}{4} \big(
        -19+140 z-76 z^2\big) \log ^2(z)
+4 \big(
        3-6 z+10 z^2\big) \log (1-z) \log ^2(z)
+\frac{1}{6} \big(
        -9+18 z-52 z^2\big) \log ^3(z)
\nonumber\\&        
- 4 (1+z) (1+3 z) \log (z) \log (1+z)
- \big(
        13+16 z-28 z^2\big) \text{Li}_2(1-z)
-2 \big(
        1-2 z+26 z^2\big) \log (1-z) \text{Li}_2(1-z)
\nonumber\\&        
- 4 (1+z) (1+3 z) \text{Li}_2(-z)
+6 \big(
        -1+2 z+6 z^2\big) \text{Li}_3(1-z)
-2 \big(
        7-14 z+34 z^2\big) \text{S}_{1,2}(1-z)
\nonumber\\&        
+2 \big(
        5-16 z+6 z^2\big) \zeta_2
-8 \big(
        1-2 z+2 z^2\big) \log (1-z) \zeta_2
+2 \big(
        19-38 z+50 z^2\big) \zeta_3
-\frac{1}{2} \big(
        35-301 z+214 z^2\big) \log (z)
\nonumber\\&        
+8 \big(
        1-2 z+6 z^2\big) \zeta_2 \log (z)
-2 (-1+2 z) \log (z) \text{Li}_2(1-z)
-16 z^2 \log (z) \text{Li}_2(-z)
+32 z^2 \text{Li}_3(-z)
\Big\}
\nonumber\\&
+ C_A \Big\{
\frac{1}{54 z} \big( -208+1185 z-2598 z^2+1513 z^3 \big)
+\frac{1}{9 z} \big(
        16-228 z+57 z^2+182 z^3\big) \log (1-z)
\nonumber\\&        
-\frac{1}{3 z} (-1+z) \big(
        16+z+145 z^2\big) \log ^2(1-z)
+\frac{13}{3} \big(
        1-2 z+2 z^2\big) \log ^3(1-z)
\nonumber\\&        
+2 \big(
        1-28 z+62 z^2\big) \log (1-z) \log (z)
+2 \big(
        1+22 z-6 z^2\big) \log ^2(1-z) \log (z)
+\frac{1}{6} \big(
        -3+108 z-292 z^2\big) \log ^2(z)
\nonumber\\&        
+2 \big(
        -3-14 z+2 z^2\big) \log (1-z) \log ^2(z)
+2 (1+z) (3+5 z) \log (z) \log (1+z)
\nonumber\\&
-8 \big(
        1+2 z+2 z^2\big) \log (1-z) \log (z) \log (1+z)
+6 \big(
        1+2 z+2 z^2\big) \log ^2(z) \log (1+z)
\nonumber\\&        
+\frac{2}{3 z} \big(
        16-12 z+48 z^2+53 z^3\big) \text{Li}_2(1-z)
+2 \big(
        13+22 z+10 z^2\big) \log (1-z) \text{Li}_2(1-z)
\nonumber\\&        
-8 (-1+z)^2 \log (z) \text{Li}_2(1-z)
+2 (1+z) (3+5 z) \text{Li}_2(-z)
-8 \big(
        1+2 z+2 z^2\big) \log (1-z) \text{Li}_2(-z)
\nonumber\\&        
+8 \big(
        1+2 z+2 z^2\big) \log (z) \text{Li}_2(-z)
-4 \big(
        7+18 z+6 z^2\big) \text{Li}_3(1-z)
-4 \big(
        1+2 z+2 z^2\big) \text{Li}_3(-z)
\nonumber\\&        
-8 \big(
        1+2 z+2 z^2\big) \text{Li}_3\Big(
        -\frac{1-z}{1+z}\Big)
+8 \big(
        1+2 z+2 z^2\big) \text{Li}_3\Big(
        \frac{1-z}{1+z}\Big)
+\frac{8}{3 z} \big(
        -2+3 z-15 z^2+20 z^3\big) \zeta_2
\nonumber\\&        
-16 \big(
        1-z+2 z^2\big) \log (1-z) \zeta_2
-2 \big(
        1+4 z+2 z^2\big) \zeta_3
+\frac{1}{9} \big(
        102-66 z-565 z^2\big) \log (z)
+8 z (-5+2 z) \zeta_2 \log (z)
\nonumber\\&
+\frac{1}{3} (5+14 z) \log ^3(z)
+16 \big(
        1+3 z+z^2\big) \text{S}_{1,2}(1-z)
\Big\}\,.
% 
%%%%%%%%%%%%%%%%%%%%%%%%%%
\\
% % 
\Delta_{gg}^{(2,0)} &= 
2 (-1+z) (10+59 z)
-(2 (-1+z) (23+75 z) \log (1-z))
+16 (-1+z) (1+3 z) \log ^2(1-z)
\nonumber\\&
-4 \big(
        -5-16 z+4 z^2\big) \log (1-z) \log (z)
-8 (1+2 z)^2 \log ^2(1-z) \log (z)
+4 (1+2 z)^2 \log (1-z) \log ^2(z)
\nonumber\\&
-\frac{2}{3} \big(
        1+4 z+8 z^2\big) \log ^3(z)
+6 \big(
        1+2 z+2 z^2\big) \log ^2(z) \log (1+z)
-4 \big(
        1+2 z+2 z^2\big) \log (z) \log ^2(1+z)
\nonumber\\&        
+4 \big(
        -1+4 z+14 z^2\big) \text{Li}_2(1-z)
-16 (1+2 z)^2 \log (1-z) \text{Li}_2(1-z)
-4 (1+2 z)^2 \log (z) \text{Li}_2(1-z)
\nonumber\\&
+4 \big(
        3+6 z+2 z^2\big) \log (z) \text{Li}_2(-z)
-8 \big(
        1+2 z+2 z^2\big) \log (1+z) \text{Li}_2(-z)
+16 (1+2 z)^2 \text{Li}_3(1-z)
\nonumber\\&
+4 \big(
        -3-6 z+2 z^2\big) \text{Li}_3(-z)
-4 \big(
        3+18 z+14 z^2\big) \text{S12}(1-z)
-4 \big(
        -4-9 z+12 z^2\big) \zeta_2
+8 \big(
        -1-2 z+z^2\big) \zeta_3
\nonumber\\&        
+\big(
        -15-48 z+121 z^2\big) \log (z)
+8 \big(
        1+4 z+5 z^2\big) \zeta_2 \log (z)
-2 \big(
        2+15 z+4 z^2\big) \log ^2(z)
\nonumber\\&        
-4 \big(
        1+2 z+2 z^2\big) \zeta_2 \log (1+z)
+8 z \log (z) \log (1+z)
+8 z \text{Li}_2(-z)
-8 \big(
        1+2 z+2 z^2\big) \text{S}_{1,2}(-z)
\nonumber\\&
+ \frac{C_A^2}{(N^2-1)} \Big\{
\frac{1}{3} (-1+z) (1+249 z)
+\frac{2}{3} z (3+25 z) \log ^2(z)
+\frac{8}{3} z (-3+2 z) \log (z) \log (1+z)
\nonumber\\&
-6 \big(
        1+2 z+2 z^2\big) \log ^2(z) \log (1+z)
+4 \big(
        1+2 z+2 z^2\big) \log (z) \log ^2(1+z)
+\frac{8}{3} z (-3+2 z) \text{Li}_2(-z)
\nonumber\\&
-12 \big(
        1+2 z+2 z^2\big) \log (z) \text{Li}_2(-z)
+8 \big(
        1+2 z+2 z^2\big) \log (1+z) \text{Li}_2(-z)
+12 \big(
        1+2 z+2 z^2\big) \text{Li}_3(-z)
\nonumber\\&        
-4 \big(
        1-2 z+2 z^2\big) \text{S}_{1,2}(1-z)
+\frac{4}{3} z (-3+2 z) \zeta_2
+8 \big(
        1+2 z+2 z^2\big) \zeta_3
-\frac{2}{3} \big(
        -2+40 z+87 z^2\big) \log (z)
\nonumber\\&        
+4 \big(
        1+2 z+2 z^2\big) \zeta_2 \log (1+z)
+8 \big(
        1+2 z+2 z^2\big) \text{S}_{1,2}(-z)
\Big\} \,.
\end{align}

The corresponding results from the QED and QCD$\times$QED are found to be
\begin{align}
 \Delta^{(1,1)}_{b\bar{b}} &= \Delta_{b\bar{b}}^{(2,0)} \Big|_{C_F^2 \rightarrow 2 C_F e_b^2, C_A C_F \rightarrow 0, C_F n_f T_F \rightarrow 0, C_F T_F \rightarrow 0}
\\
 \Delta^{(1,1)}_{bb} &= \Delta_{bb}^{(2,0)} \Big|_{C_F^2 \rightarrow 2 C_F e_b^2, C_A C_F \rightarrow 0, C_F n_f T_F \rightarrow 0, C_F T_F \rightarrow 0}
\\
 \Delta^{(1,1)}_{ub} &= 0
\\
 \Delta^{(1,1)}_{u\bar{u}} &= 0
\\
 \Delta^{(1,1)}_{bg} &= \frac{1}{2 C_A C_F} \left[ \Big(2 C_A C_F \Delta_{bg}^{(2,0)} \Big) \Big|_{C_A C_F^2 \rightarrow C_A C_F e_b^2, C_A^2 C_F \rightarrow 0} \right]
                = \Delta_{bg}^{(2,0)} \Big|_{C_F \rightarrow e_b^2, C_A \rightarrow 0}
\\
 \Delta^{(1,1)}_{b\gamma} &= \Big(2 C_A C_F \Delta_{bg}^{(2,0)} \Big) \Big|_{C_A C_F^2 \rightarrow C_A C_F e_b^2, C_A^2 C_F \rightarrow 0}
\\
 \Delta^{(1,1)}_{g\gamma} &= \frac{1}{2 C_A C_F} \left[ \Big(4 C_A^2 C_F^2 \Delta_{gg}^{(2,0)} \Big) \Big|_{C_A^2 C_F^2 \rightarrow C_A^2 C_F e_b^2, 
                             C_A^3 C_F \rightarrow 0} \right]
 = \Big(2 C_A C_F \Delta_{gg}^{(2,0)} \Big) \Big|_{C_A C_F \rightarrow C_A e_b^2, C_A^2 \rightarrow 0}
\end{align}

Partonic cross sections contributing to pure NLO and NNLO QED corrections:
\begin{align}
 \Delta^{(0,1)}_{b\bar{b}} &= \Delta_{b\bar{b}}^{(1,0)} \Big|_{C_F \rightarrow e_b^2}
\\
 \Delta^{(0,1)}_{b\gamma} &= \Big(2 C_A C_F \Delta_{bg}^{(1,0)} \Big) \Big|_{C_A C_F \rightarrow C_A e_b^2}
\\
 \Delta^{(0,2)}_{b\bar{b}} &= \Delta_{b\bar{b}}^{(2,0)} \Big|_{C_F^2 \rightarrow e_b^4, C_A C_F \rightarrow 0, 
               C_F n_f T_F \rightarrow e_b^2 N ( \sum_q e_q^2 ), C_F T_F \rightarrow N e_b^4}
\\
 \Delta^{(0,2)}_{bb} &= \Delta_{bb}^{(2,0)} \Big|_{C_F^2 \rightarrow e_b^4, C_A C_F \rightarrow 0, 
               C_F n_f T_F \rightarrow e_b^2 N ( \sum_q e_q^2 ), C_F T_F \rightarrow N e_b^4}
\\
 \Delta^{(0,2)}_{ub} &= \Delta_{ub}^{(2,0)} \Big|_{C_F T_F \rightarrow N e_b^2 e_u^2}
\\
 \Delta^{(0,2)}_{u\bar{u}} &= \Delta_{u\bar{u}}^{(2,0)} \Big|_{C_F T_F \rightarrow N e_b^2 e_u^2}
\\
 \Delta^{(0,2)}_{b\gamma} &= \Big(2 C_A C_F \Delta_{bg}^{(2,0)} \Big) \Big|_{C_A C_F^2 \rightarrow C_A e_b^4, C_A^2 C_F \rightarrow 0}
\\
 \Delta^{(0,2)}_{\gamma\gamma} &=  \Big(4 C_A^2 C_F^2 \Delta_{gg}^{(2,0)} \Big) \Big|_{C_A^2 C_F^2 \rightarrow C_A^2 e_b^4, 
                             C_A^3 C_F \rightarrow 0} 
\end{align}
% where
% 
The constants $\zeta_i = \sum_{k=1}^{\infty} \frac{1}{k^i}$, $k \in {\mathbb{N}}$  
denote the Riemann's $\zeta$-functions, e.g.,   
%\begin{align}
$\zeta_2 = 1.64493406684822643647\ldots \,$
and
$\zeta_3 = 1.20205690315959428540\ldots$.
%\end{align}
The Spence functions Li$_2(x)$ and Li$_3(x)$ are defined by
\begin{align}
 \text{Li}_2 (x) &= \sum_{k=1}^{\infty} \frac{x^k}{k^2} = - \int_0^x \frac{\log (1-t)}{t} dt \,,
 \nonumber\\
 \text{Li}_3 (x) &= \sum_{k=1}^{\infty} \frac{x^k}{k^3} = \int_0^x \frac{\text{Li}_2 (t)}{t} dt \,,
\end{align}
and the Nielson function S$_{1,2}(x)$ is given by
\begin{align}
 \text{S}_{1,2} (x) &= \frac{1}{2} \int_0^1 {dt \over t} \log^2(1-t x) \,.
\end{align}
\end{widetext}

%-------------------------------------------------
% \begin{thebibliography}{99}

\providecommand{\href}[2]{#2}\begingroup\raggedright\endgroup

% \end{thebibliography}

\end{document}